\documentclass{kapedbk} 
\usepackage[dvips]{graphicx}
\upperandlowercase
\kluwerbib 
\normallatexbib 
\begin{document}
\articletitle{Precursors of catastrophic ~\\
 failures }

\author{Srutarshi Pradhan\altaffilmark{1} and  Bikas K. Chakrabarti\altaffilmark{2}}
\vskip.3in
\noindent {Saha Institute of Nuclear Physics, 1/AF Bidhan Nagar, 
Kolkata 700064, India}
\altaffiltext{1}{e-mail: spradhan@cmp.saha.ernet.in}
\altaffiltext{2}{e-mail: bikas@cmp.saha.ernet.in}




\begin{abstract}
\noindent We review here briefly the nature of precursors of global failures 
in three different kinds of many-body dynamical systems. First, we consider 
the lattice models of self-organised criticality in sandpiles and investigate 
numerically the effect 
of pulsed perturbations to the systems prior to reaching their respective 
critical points.
We consider next, the random strength fiber bundle models, under global load 
sharing approximation, and derive analytically the partial failure response 
behavior at loading level less than its global failure or critical point. 
Finally, we consider the two-fractal overlap model of earthquake and analyse 
numerically the overlap time series data as one fractal moves over the other 
with uniform velocity. The precursors of global or major failure in all three 
cases are shown to be very well characterized and prominent.   
\end{abstract}



\vskip .2in







\section{Introduction}

\noindent A major failure of a solid or a dynamic catastrophe can often
be viewed as a phase transition from an unbroken or non-chaotic phase
to a broken or chaotic phase. Some obvious correlations existing in
the system before the failure, are lost in the failure process. However,
the equivalence can be made precise and in fact the critical behavior
goes to the bone of any failure or catastrophic phenomena. 

Several dynamical models of cooperative failure dynamics 
\cite{books} are now well-studied and their criticality at the global failure 
point are well established. Here, we have reviewed some of the numerical 
studies of precursor 
behavior in two self-organising dynamical \cite{books} models as one 
approaches the self-organised critical (SOC) point
of the avalanches in the lattice sandpile  models. The critical behavior of 
a random fiber bundle model \cite{Peirce} of failure or fracture of a solid 
under global load sharing is now very precisely demonstrated. The dynamics
of failure also become critically slow there and all these have been 
demonstrated analytically. The precursors of the global failure in the fiber 
bundle models can therefore be discussed analytically.  The resulting 
universality of the surface roughness
in any such fracture process has also been well documented and analysed.
The two-fractal overlap model \cite{BS99} is a very recent modeling approach 
of earthquake dynamics. The time series of overlap magnitudes obtained, when 
one fractal slides over the other with uniform velocity, represents the model 
seismic activity variations. This time series analysis suggests   
some precursors of large events in this model.

\section{Precursors in SOC models of sandpile}

\noindent A `pile' of dry sand is a unique example 
SOC system in nature.  A growing sandpile gradually comes to a `quasi-stable' 
state through its self-organising dynamics. This `quasi-stable' state is 
called the critical state of the system as the system exhibits power law 
behavior there. Because of avalanches of all sizes, this critical point in 
the pile is also a catastrophic one.  The dynamics of growing sandpiles are 
successfully modeled  by the lattice model \cite{BTW} of Bak, Tang and 
Wisenfeld (BTW) in 1987.  
A stochastic version of sand pile model has been introduced 
by Manna \cite {Manna} which also shows SOC, although it belongs to a 
different universality
class. Both the models have been studied extensively at their criticality. 
Here, we study the sub-critical behavior of both the models and look for the 
precursors of the critical state.


\subsubsection{BTW Model}
Let us consider a square lattice of size \( L\times L \).
At each lattice site \( (i,j) \), there is an integer variable \( h_{i,j} \)
which represents the height of the sand column at that site. A unit
of height (one sand grain) is added at a randomly chosen site at each
time step and the system evolves in discrete time. The dynamics starts
as soon as any site \( (i,j) \) has got a height equal to the threshold
value (\( h_{th} \)= \( 4 \)): the site topples, i.e., \( h_{i,j} \)
becomes zero there, and the heights of the four neighboring sites
increase by one unit \begin{equation}
\label{00}
h_{i,j}\rightarrow h_{i,j}-4,h_{i\pm 1,j}\rightarrow h_{i\pm 1,j}+1,h_{i,j\pm 1}\rightarrow h_{i,j\pm 1}+1.
\end{equation}
 If, due to this toppling at site \( (i,j) \), any neighboring site
become unstable (its height reaches the threshold value), they in
turn follow the same dynamics. The process continues till all sites
become stable (\( h_{i,j}< \) \( h_{th} \) for all \( (i,j) \)).
When toppling occurs at the boundary of the lattice (four nearest
neighbors are not available), extra heights get off the lattice and
are removed from the system.

With a very slow but steady rate of addition of unit height (sand
grain) at random sites of the lattice, the avalanches get correlated
over longer and longer ranges and the average height (\( h_{av} \))
of the system grows with time. Gradually the correlation length (\( \xi  \))
becomes of the order the system size \( L \). Here, on average, the
additional height units start leaving the system as the system approaches
toward a critical average height \( h_{c}(L) \) and the average height
\( h_{av} \) remains stable there (see Fig. 1(a)). Also the system becomes 
critical here (for $L$ $\rightarrow$ $\infty$) as the distributions of the avalanche sizes and the 
corresponding life times follow robust power laws \cite{BTW}. 

\vspace{0.3cm}
\resizebox*{6cm}{6.5cm}{\rotatebox{-90}{\includegraphics{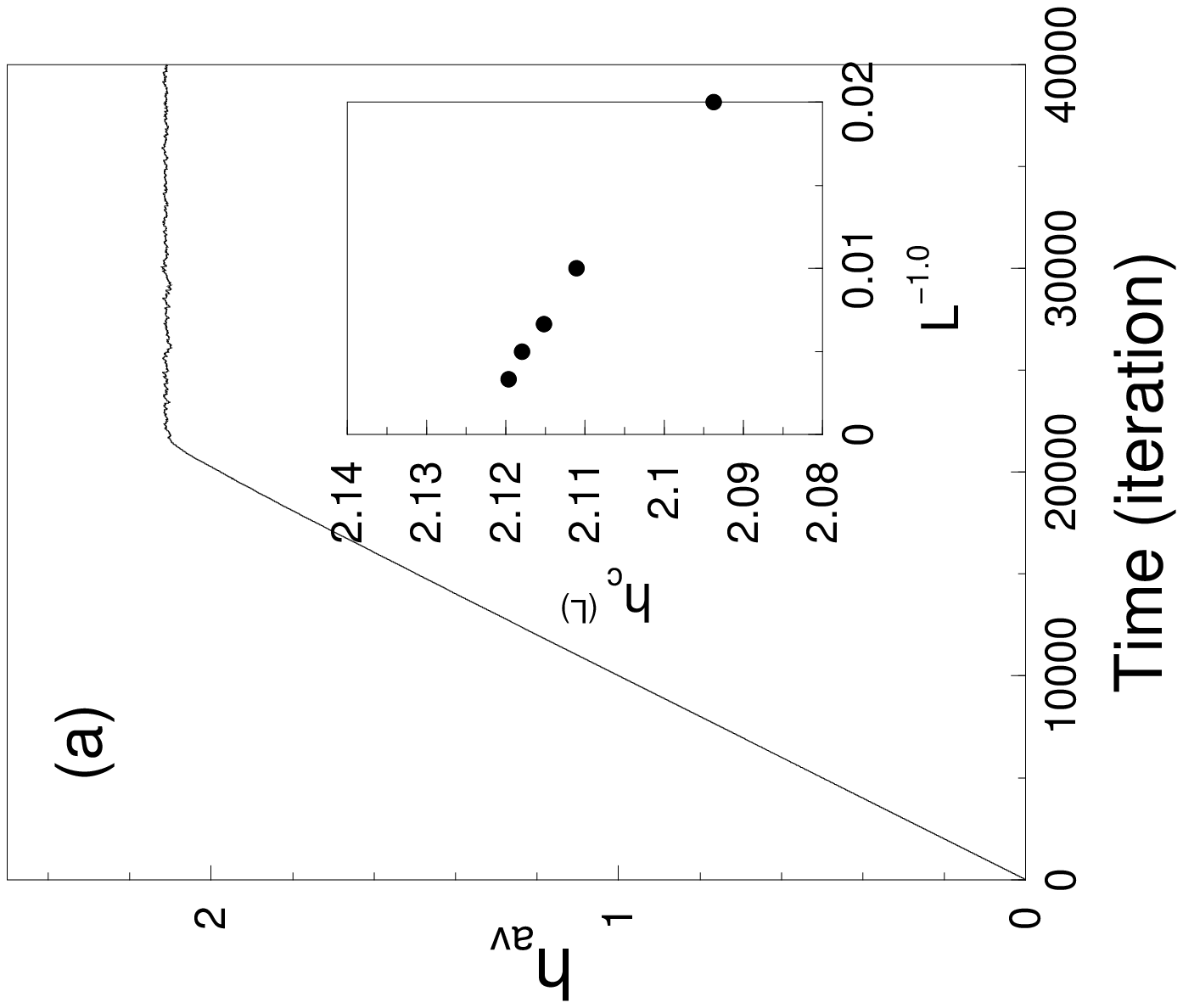}}}
\resizebox*{6cm}{6.5cm}{\rotatebox{-90}{\includegraphics{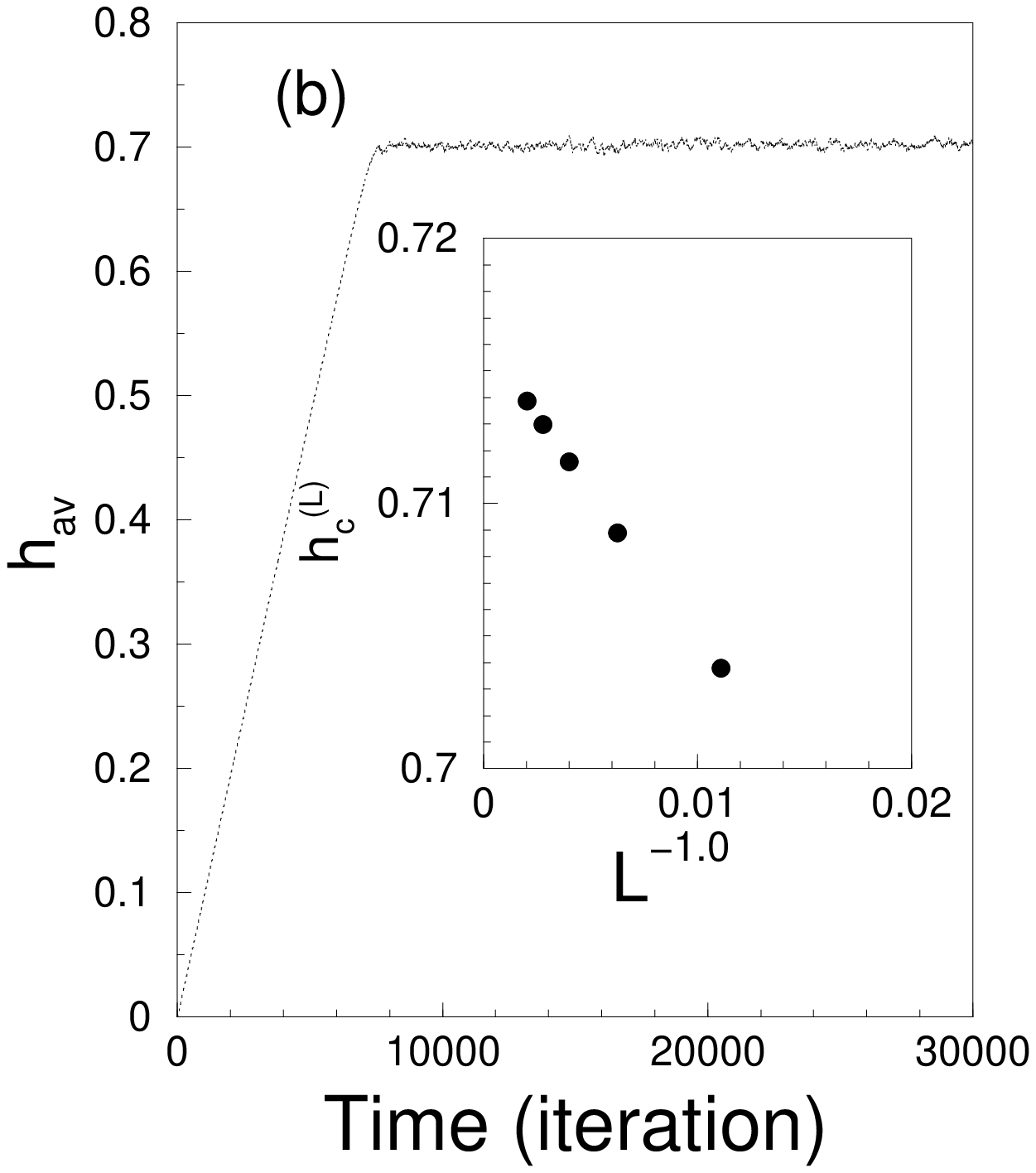}}}
\vspace{0.3cm}

\noindent\textbf{\footnotesize \it Figure 1.} {\footnotesize The growth of 
average height \( h_{av} \) against the
number of iterations of adding unit heights \( (L=100) \). Eventually 
\( h_{av} \) settles at \( h_{c}(L) \). In the
inset, we show the finite size behavior of the critical height \( h_{c}(L) \),
obtained from simulation results for different \( L \). (a) For the BTW model; 
(b) for the Manna model.} {\footnotesize \par}
\vskip.2in
Here, a finite size scaling fit \( h_{c}(L)=h_{c}(\infty )+{\textrm{C}}L^{-1/\nu } \)
(obtained by setting \( \xi  \) \( \sim  \) \( \mid h_{c}(L)-h_{c}(\infty )\mid ^{-\nu }=L \)),
where C is a constant, with \( \nu \simeq 1.0 \) gives \( h_{c}\equiv h_{c}(\infty )\simeq 2.124 \)
(see inset of Fig. 1(a)). Similar finite size scaling fit 
with \( \nu =1.0 \)
gave \( h_{c}(\infty )\simeq 2.124 \) in earlier large scale simulations 
\cite{SB01}.
\subsubsection{Manna Model}

\noindent BTW model is a deterministic one. Manna proposed the stochastic
sand-pile model \cite{Manna} by introducing randomness in the dynamics
of sandpile growth. Here, the critical height is \( 2 \). Therefore
at each toppling the rejected two grains choose their host among the
four available neighbors randomly with equal probability. After constant
adding of sand grains, the system ultimately settles at a critical
state. 

We consider now the Manna model on a square lattice of size
\( L\times L \), where the sites can be either empty or occupied
with unit height i.e., the height variables can have binary states
\( h_{i,j}=1 \) or \( h_{i,j}=0 \). A site is chosen randomly and
one height is added at that site. If the site is initially empty,
it gets occupied: \begin{equation}
\label{pp}
h_{i,j}\rightarrow h_{i,j}+1,
\end{equation}
 If the chosen site is previously occupied then a toppling or `hard
core interaction' rejects both the heights from that site: \begin{equation}
\label{qq}
h_{i,j}\rightarrow h_{i,j}-2,
\end{equation}
 and each of these two rejected heights stochastically chooses its
host among the \( 4 \) neighbors of the toppled site. The toppling
can happen in chains if any chosen neighbor was previously occupied
and thus cascades are created. After the system attains stable state
(dynamics stopped), a new site is chosen randomly and unit height
is added to it. Thus the system evolves in discrete time steps. Here
again the boundary is assumed to be completely absorbing so that heights
can leave the system due to the toppling at the boundary.

With a slow rate of addition of heights (sand grains) at random sites,
initially the average height of the system grows with time and soon
the system approaches toward a critical average height \( h_{c} \)
(see Fig. 1(b)). 
Here also the critical average height \( h_{c} \) has a finite size
dependence and a similar finite size scaling fit \( h_{c}(L)=h_{c}
(\infty )+{\textrm{C}}L^{-1/\nu } \)
gives \( \nu \simeq 1.0 \) and \( h_{c}\equiv h_{c}(\infty )\simeq 0.716 \)
(see inset of Fig. 1(b)). This is close to an earlier estimate \( h_{c}\simeq 
0.71695 \) \cite{SB01}, made in a somewhat different version of the model. The
avalanche size distribution has got power laws similar to the BTW
model, at this self-organised critical state at \( h_{av}=h_{c} \).
However the exponents seem to be different \cite{Manna} for this 
stochastic model, compared to those of BTW model.

\subsection{Precursors of the SOC point:}
\subsubsection{In the BTW model}

\noindent At an average
height \( h_{av} \) (\( <h_{c}(L) \)), when all sites of the system have 
become stable
(dynamics have stopped), a fixed number of height units \( h_{p}=4 \)
(pulse of sand grains) is added at any central point of the system.
Just after this addition, the local dynamics starts and it takes a
finite time or iterations (\( \tau \)) to return back to the stable state 
(\( h_{i,j}<4 \) for all \( (i,j) \)) after several toppling events 
\( \Delta \); and the disturbance spreads over a length \( \xi \), the 
correlation length of the system \cite{AC96,SB01}.  
All these three  parameters are seen \cite {SB01} to diverge as \( h_{av} \)
approaches the critical height \( h_{c} \) from below
following  power laws \( \tau \sim (h_{c}-h_{av})^{-\gamma } \),
where \( \gamma \cong 1.2 \); 
 \( \Delta  \) \( \sim (h_{c}-h_{av}) \)\( ^{-\delta } \), where \( \delta 
\cong 2.0 \); \( \xi  \) \( \sim  \) \( (h_{c}-h_{av})^{-\nu } \), where 
\( \nu \cong 1.0 \) (see Fig. 2). 

\resizebox*{5.5cm}{5.5cm}{\rotatebox{-90}{\includegraphics{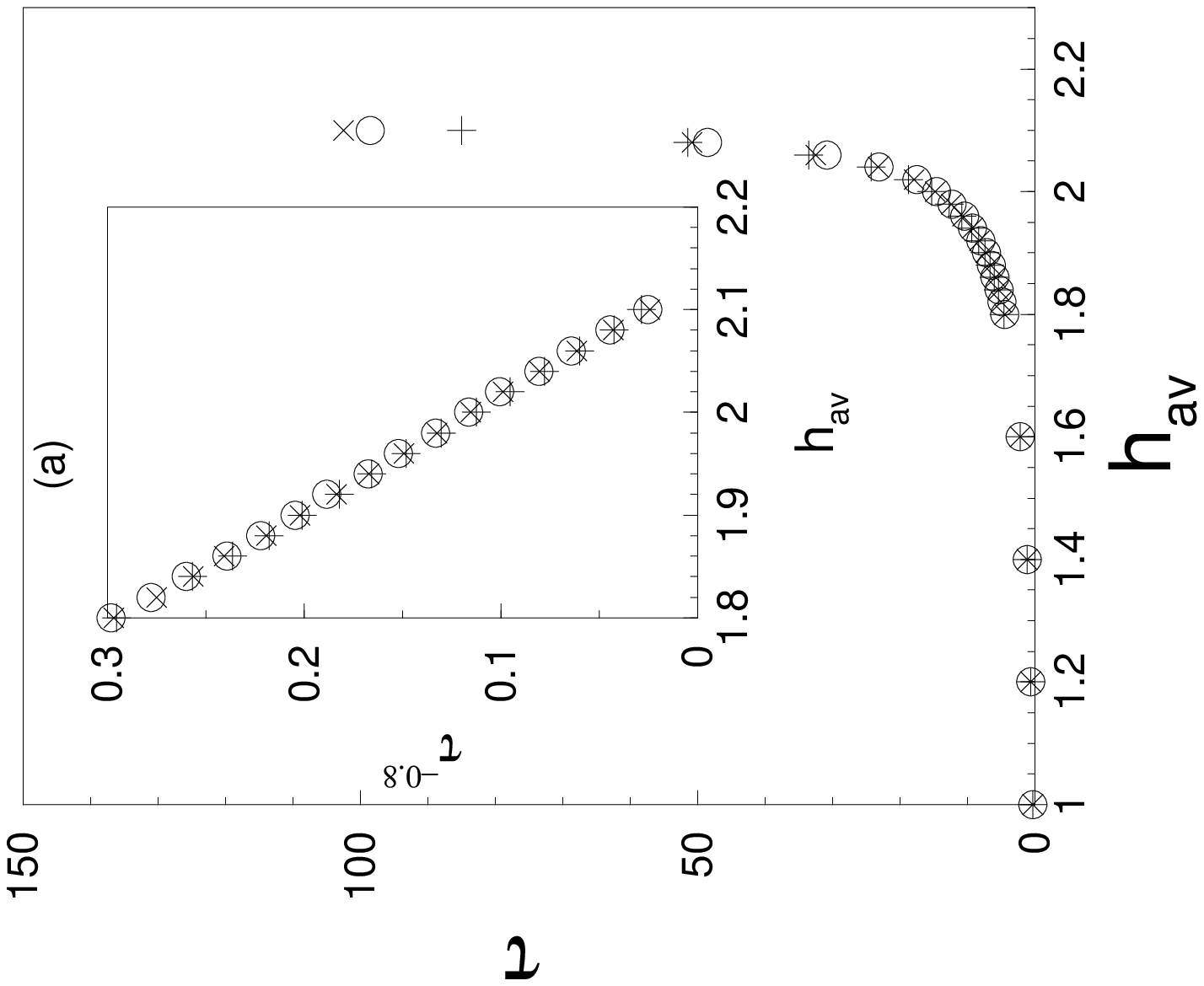}}} 
\resizebox*{5.5cm}{5.5cm}{\rotatebox{-90}{\includegraphics{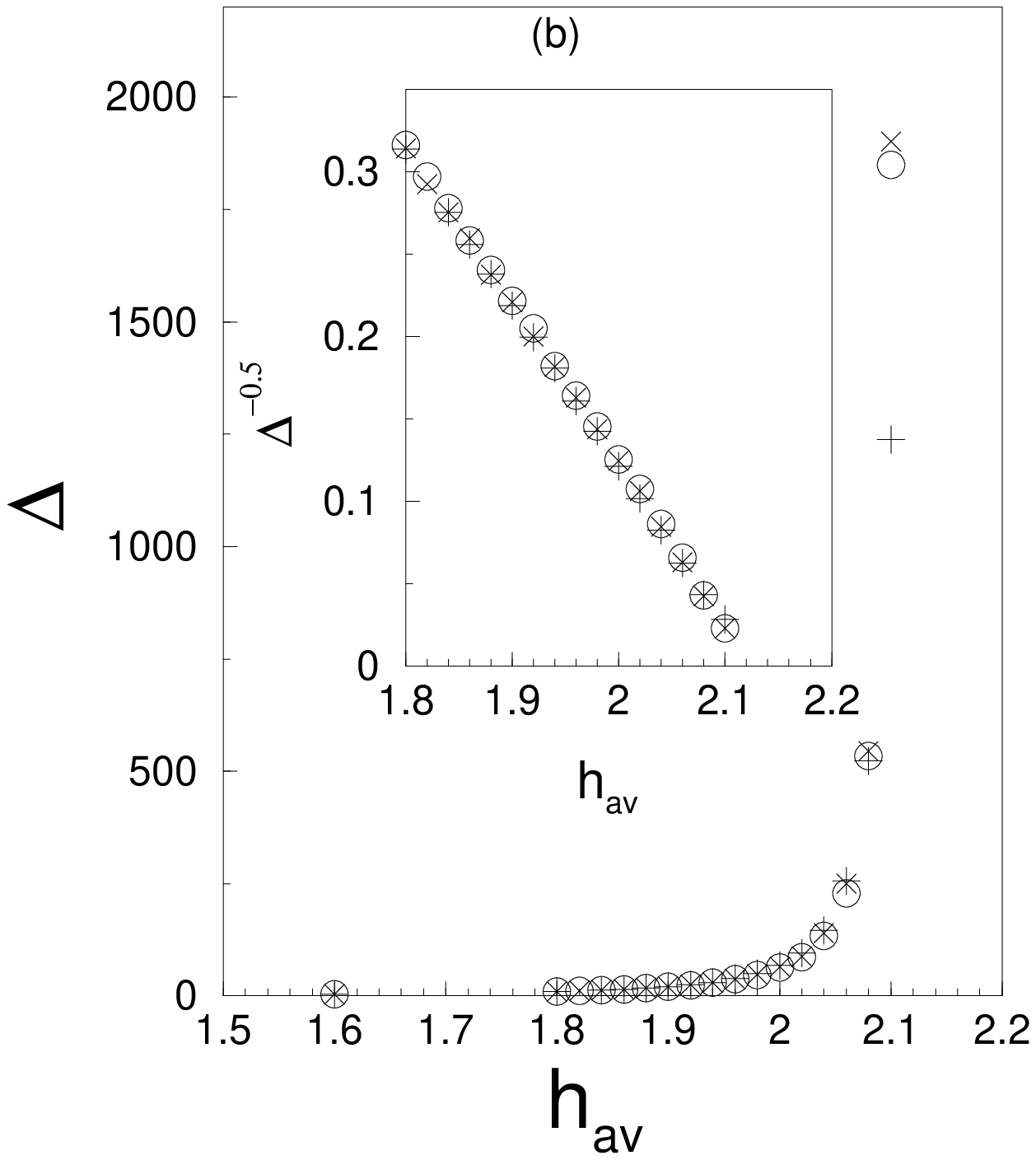}}} 

\vspace{0.1cm}
{\centering \resizebox*{5.5cm}{5.5cm}{\rotatebox{-90}{\includegraphics{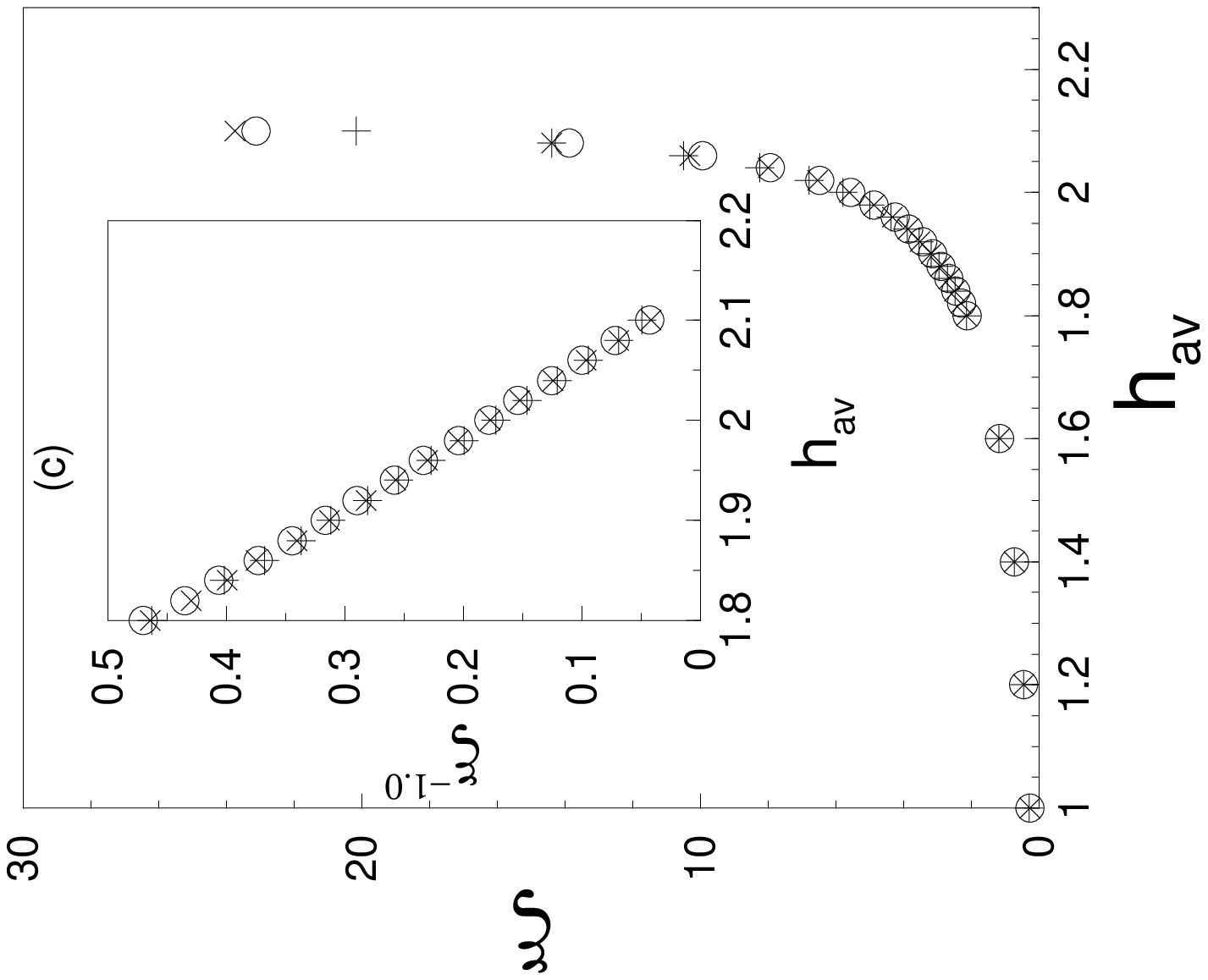}}} \par}
\vspace{0.1cm}

\noindent\textbf{\footnotesize \it Figure 2.} {\footnotesize The variations 
of the precursors with \( h_{av} \) (\( <h_{c}(L) \)) in the BTW model
for different system sizes: \( L=100 \) (plus) \( L=200 \) (cross)
and \( L=300 \) (open circle). (a) For relaxation time \( \tau  \);
in the inset \( \tau ^{-0.8} \) is plotted against \( h_{av} \).
(b) For the total number of topplings \( \Delta  \); inset shows
\( \Delta ^{-0.5} \) versus \( h_{av} \) plot. (c) For the correlation
length \( \xi  \); in the inset, \( \xi ^{-1.0} \) is plotted against
\( h_{av} \).}{\footnotesize \par}

\subsubsection{In the Manna Model}

\noindent The precursor parameters in Manna model have been measured 
using similar method as in BTW model. 
Here the pulse of height is \( h_{p}=2 \) and is added to any arbitrary 
central site. 
We get \cite{SB01} exactly  similar power law behaviors for all the parameters 
as the critical point is approached from below: 
 \( \tau \sim (h_{c}-h_{av})^{-\gamma } \),
where \( \gamma \cong 1.2 \); 
 \( \Delta  \) \( \sim (h_{c}-h_{av}) \)\( ^{-\delta } \), where \( \delta 
\cong 2.0 \); 
 \( \xi  \) \( \sim  \) \( (h_{c}-h_{av})^{-\nu } \), where \( \nu \cong 1.0 \)
(see Fig. 3). 
\vspace{0.1cm}

\resizebox*{5cm}{5cm}{\rotatebox{-90}{\includegraphics{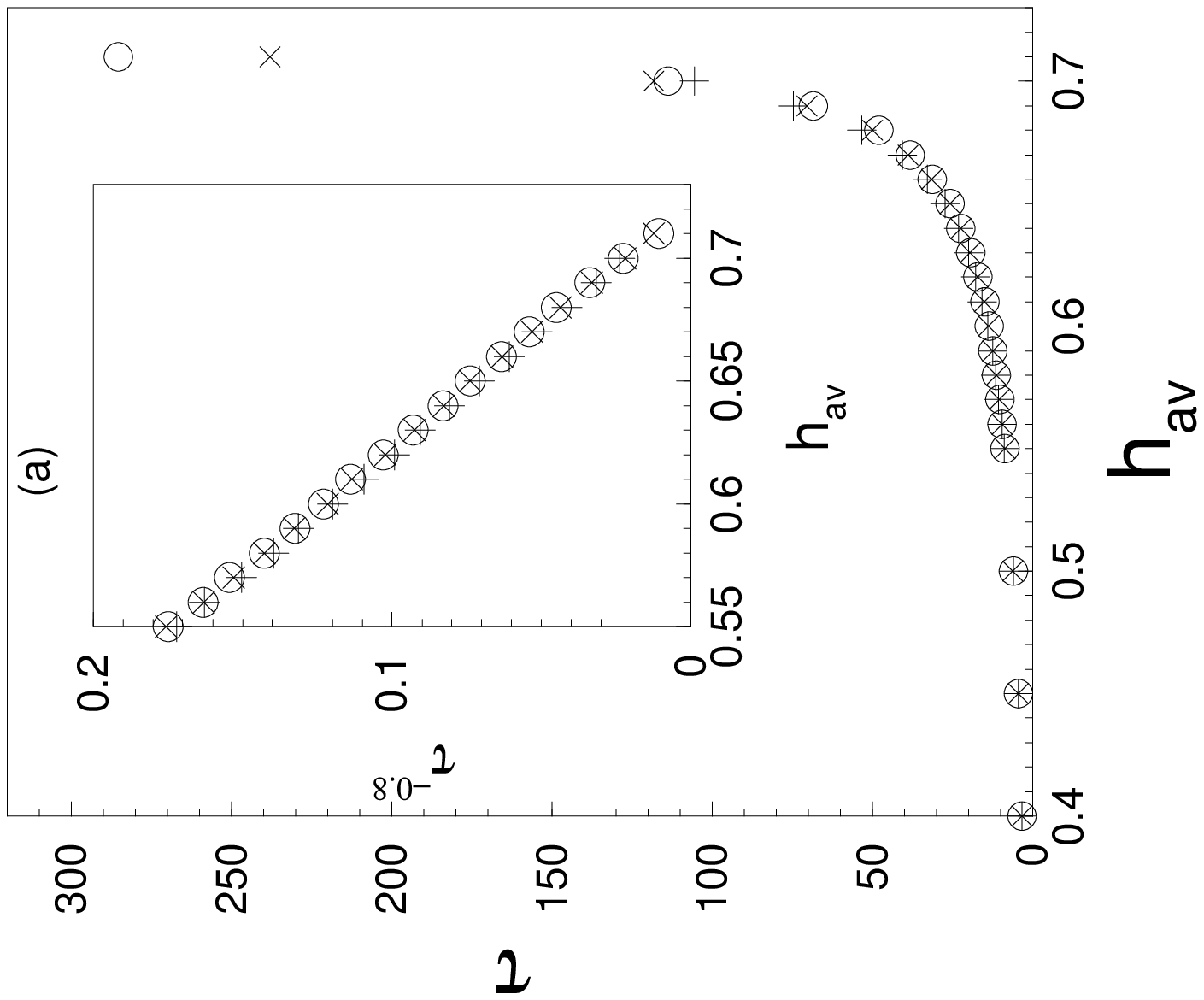}}} 
\resizebox*{5cm}{5cm}{\rotatebox{-90}{\includegraphics{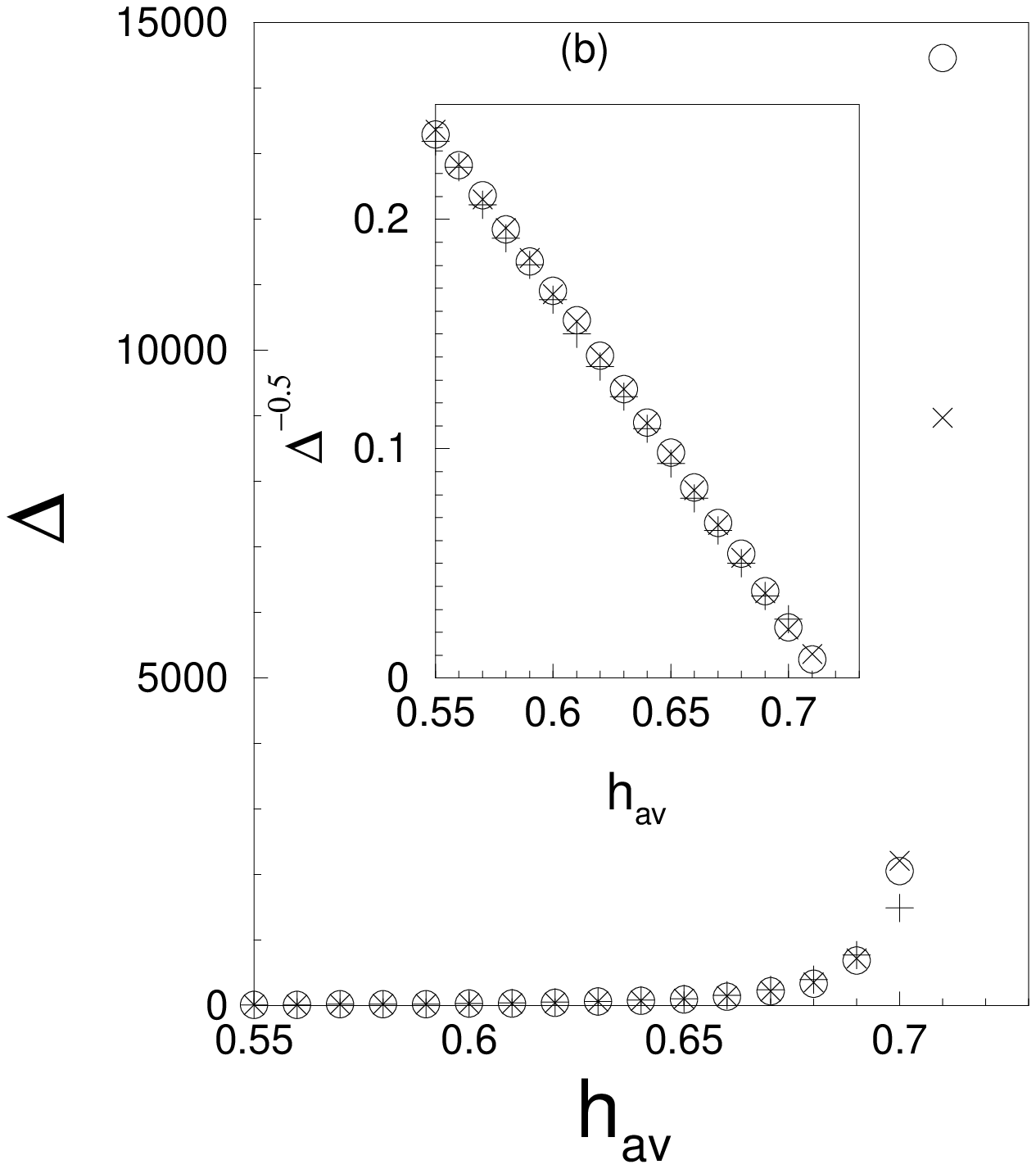}}} 

\vspace{0.1cm}
{\centering \resizebox*{5cm}{5cm}{\rotatebox{-90}{\includegraphics{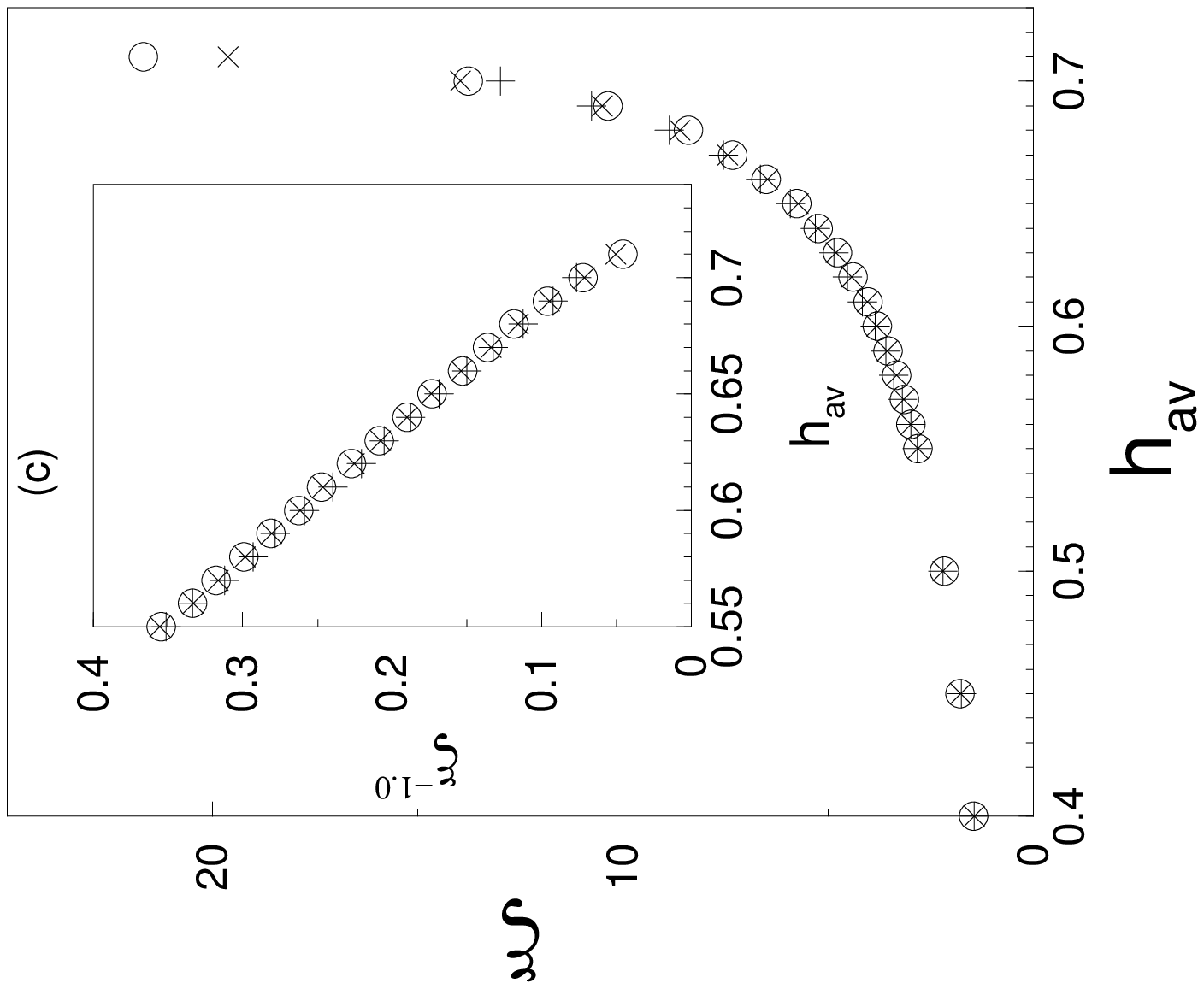}}} \par}
\vspace{0.1cm}

\noindent\textbf{\footnotesize \it Figure 3.} {\footnotesize The variations 
of the precursors with \( h_{av} \) (\( <h_{c}(L) \)) in the Manna model
for different system sizes: \( L=100 \) (plus) \( L=200 \) (cross)
and \( L=300 \) (open circle). (a) For relaxation time \( \tau  \);
in the inset \( \tau ^{-0.8} \) is plotted against \( h_{av} \).
(b) For the total number of topplings \( \Delta  \); inset shows
\( \Delta ^{-0.5} \) versus \( h_{av} \) plot. (c) For the correlation
length \( \xi  \); in the inset, \( \xi ^{-1.0} \) is plotted against
\( h_{av} \).}{\footnotesize \par}

The Monte Carlo studies showed that  these response  parameters like the 
relaxation time \( \tau \), 
size of the damage \( \Delta \) and its radial size \( \xi \), all tend to 
diverge, following the above mentioned robust power laws, in both the models. 
Precise knowledge of these power laws can therefore help estimating the 
critical or catastrophic point ($h_{c}$) by extrapolating the inverse power 
of these quantities for informations at  $h_{av}$ ($< h_{c}$) 
(see insets of Figs. 2 and 3). Similar critical slowing down phenomena should 
be of interest in the context of magnetohydrodynamics of plasma \cite{plasma}.  

\section{Precursors of fracture-failure in fiber bundles}

\noindent 
The fiber bundle model was introduced by Peirce \cite{Peirce} in
the context of testing the strength of cotton yarns. 
Fiber bundles
are of two classes with respect to the time dependence of fiber strength:
The `static' bundles contain fibers whose strengths are
independent of time, where as the `dynamic' bundles  are
assumed to have time dependent elements to capture the creep rupture
and fatigue behaviors. According to the load sharing rule,  fiber
bundles are being classified into two groups: Global load-sharing
(democratic) bundles and local load-sharing  bundles. In democratic
bundles intact fibers bear the applied load equally and in local load-sharing
bundles the terminal load of the failed fiber is given equally to
all the intact neighbors. With steadily increasing load, the fiber
bundles approach their respective failure point obeying a dynamics
determined by the load sharing rule. The phase transition \cite{SBP02}
and dynamic critical behavior of the fracture process in such democratic
bundles has been established through recursive formulation \cite{Peirce,SBP02}
of the failure dynamics. The exact solutions
of the recursion relations in the global load-sharing case suggest universal 
values of the exponents involved.

We discuss here the failure of a static fiber bundles under global
load sharing (democratic bundles) with steadily increasing load on
the bundles. We show analytically the variation of associated precursor
parameters with the applied stress which help to estimate the failure
point accurately.





The model assumes equal load sharing, i.e., the intact
fibers share the applied load equally. The strength of each of the
fibers in the bundle is determined by the stress value (\( \sigma _{th} \))
it can bear, and beyond which it fails. The strength of the fibers
are taken from a randomly distributed normalised density \( \rho (\sigma _{th}) \)
within the interval \( 0 \) and \( 1 \) such that 
\begin{equation}
\int _{0}^{1}\rho (\sigma _{th})d\sigma _{th}=1.
\end{equation}
 The global load sharing assumption neglects `local' fluctuations in stress 
(and its redistribution) and renders the model as a mean-field
one.


\subsubsection{Breaking dynamics of the democratic bundles}

\vskip.1in

\noindent The breaking dynamics starts when an initial stress \( \sigma  \)
(load per fiber) is applied on the bundle. The fibers having strength
less than \( \sigma  \) fail instantly. Due to this rupture, total
number of intact fibers decreases and now these intact fibers have
to bear the applied load on the bundle. Hence effective stress on
the fibers increases and this compels some more fibers to break. These
two sequential operations, the stress redistribution and further breaking
of fibers continue till an equilibrium is reached, where either the
surviving fibers are strong enough to bear the applied load on the
bundle or all fibers fail. 

This breaking dynamics can be represented by recursion
relations in discrete time steps. Let \( U_{t} \) be the fraction
of fibers in the initial bundle that survive after time step \( t \),
where time step indicates the number of (elemental) stress redistributions.
Then the redistributed load per fiber after \( t \) time step becomes
\begin{equation}
\label{may14-1}
\sigma _{t}=\frac{\sigma }{U_{t}};
\end{equation}

\noindent and after \( t+1 \) time steps the surviving fraction of
fiber is \begin{equation}
\label{may14-2}
U_{t+1}=1-P(\sigma _{t});
\end{equation}

\noindent where \( P(\sigma _{t}) \) is the cumulative probability
of corresponding density distribution \( \rho (\sigma _{th}) \):
\begin{equation}
\label{may14-3}
P(\sigma _{t})=\int ^{\sigma _{t}}_{0}\rho (\sigma _{th})d\sigma _{th}.
\end{equation}
Using now Eqns. (1.5) and (1.6) we can write the recursion relations
which show how \( \sigma _{t} \) and \( U_{t} \) evolve in discrete
time: \begin{equation}
\label{may14-4}
\sigma _{t+1}=\frac{\sigma }{1-P(\sigma _{t})};\sigma _{0}=\sigma 
\end{equation}

\noindent and \begin{equation}
\label{may14-5}
U_{t+1}=1-P(\sigma /U_{t});U_{0}=1.
\end{equation}

The recursion relations (1.5) and (1.6) represent the basic dynamics of
failure in global load sharing models. At the equilibrium or steady
state \( U_{t+1}=U_{t}\equiv U^{*} \) and \( \sigma _{t+1}=\sigma _{t}\equiv \sigma ^{*} \).
This corresponds to a fixed point of the recursive dynamics. Eqn. (1.6)
can be solved at the fixed point for some particular distribution
of \( \rho (\sigma _{th}) \) and these solutions near \( U^{*} \)
(or \( \sigma ^{*} \)) give the detail features of the failure dynamics
of the bundle. 

\vskip.2in

\subsection{Precursors of global failure:}
\subsubsection{For uniform distribution of fiber strength}
\noindent We choose first the uniform density of fiber strength distribution
to solve the recursive failure dynamics of democratic bundle. Here,
the cumulative probability becomes \begin{equation}
\label{may20-1}
P(\sigma _{t})=\int ^{\sigma _{t}}_{0}\rho (\sigma _{th})d\sigma _{th}=\int ^{\sigma _{t}}_{0}d\sigma _{th}=\sigma _{t}.
\end{equation}
Therefore \( U_{t} \) follows a simple recursion relation (following
Eqn. (1.6)) \begin{equation}
\label{qrw}
U_{t+1}=1-\frac{\sigma }{U_{t}}.
\end{equation}

\resizebox*{4cm}{4cm}{\includegraphics{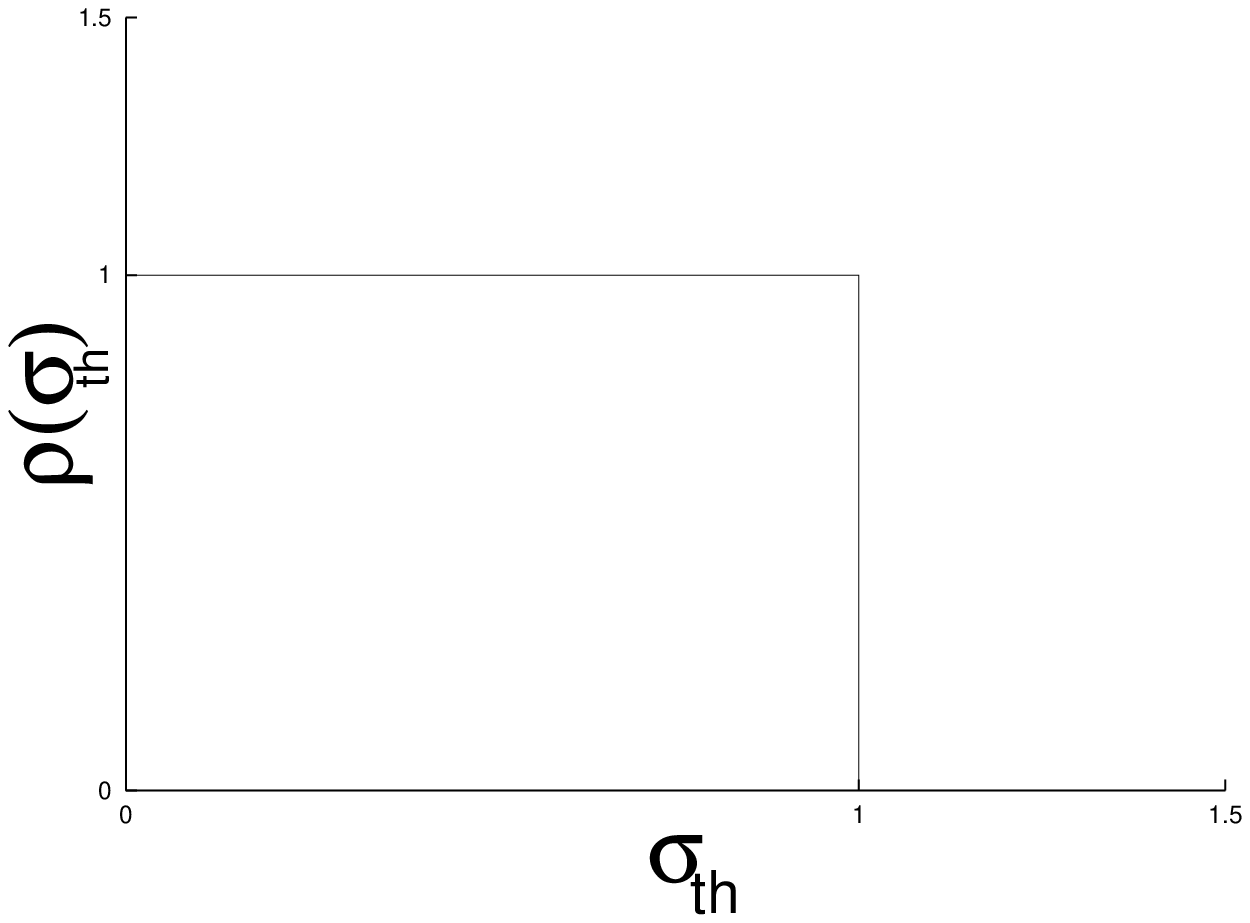}} 
\resizebox*{4cm}{4cm}{\includegraphics{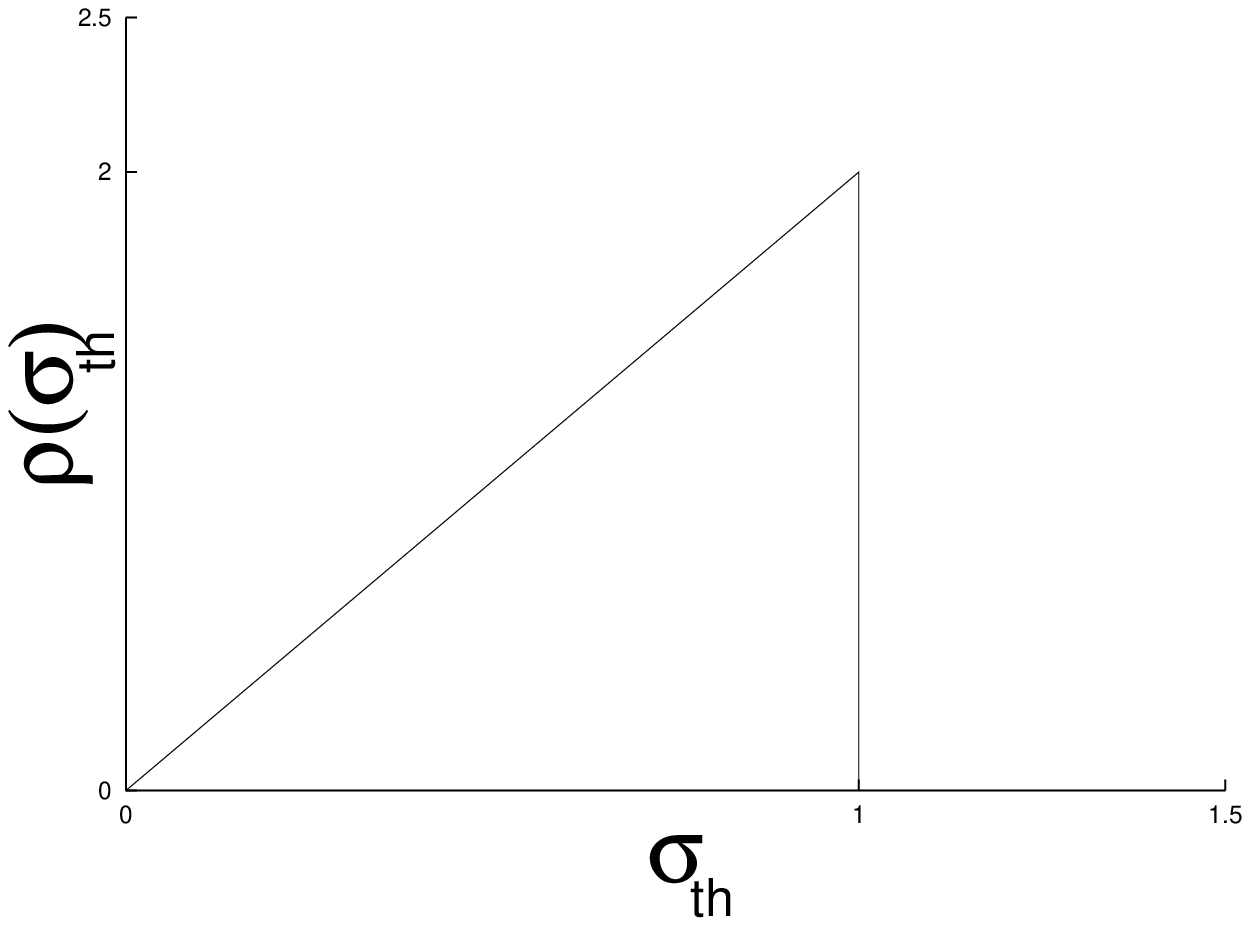}} 
\resizebox*{4cm}{4cm}{\includegraphics{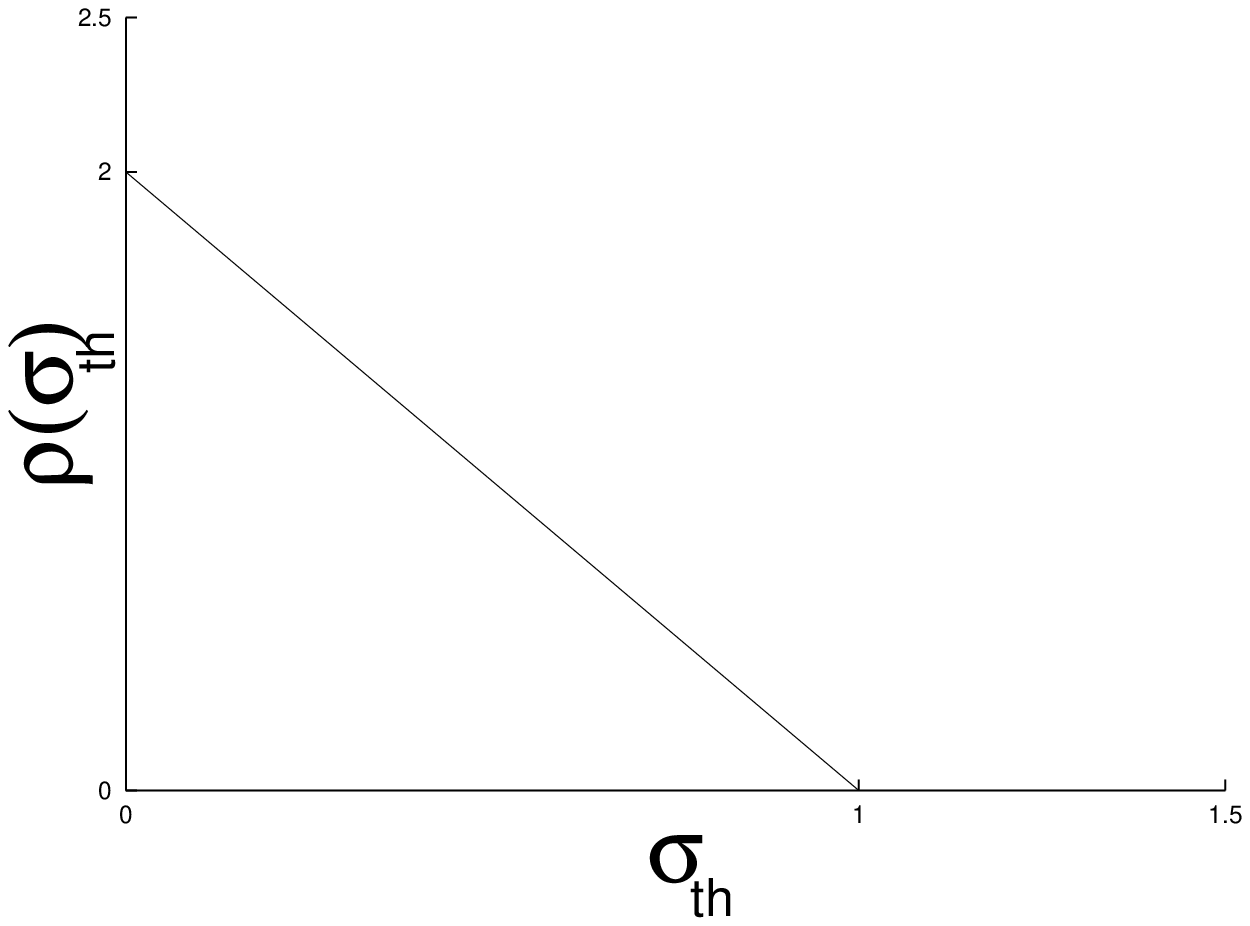}} 

\vskip.1in

{\noindent \textbf{\footnotesize \it Figure 4.} {\footnotesize Three simple
models considered here assume (a) uniform, (b) linearly increasing and (c) 
linearly decreasing density \( \rho (\sigma _{th}) \)
of the fiber strength distribution up to a cutoff strength. }\footnotesize \par}

\vskip.1in

At the equilibrium state (\( U_{t+1}=U_{t}=U^{*} \)), the above relation
takes a quadratic form of \( U^{*} \) : \begin{equation}
\label{may14-6}
U^{*^{2}}-U^{*}+\sigma =0.
\end{equation}

\noindent The solution is

\noindent \begin{equation}
\label{qq}
U^{*}(\sigma )=\frac{1}{2}\pm (\sigma _{c}-\sigma )^{1/2};\sigma _{c}=\frac{1}{4}.
\end{equation}

\noindent Here \( \sigma _{c} \) is the critical value of the initial
applied stress beyond which the bundle fails completely. The solution
with (\( + \)) sign is the stable one, whereas the one with (\( -) \)
sign gives unstable solution. The quantity
\( U^{*}(\sigma ) \) must be real valued as it has a physical meaning:
it is the fraction of the original bundle that remains intact under
a fixed applied stress \( \sigma  \) when the applied stress lies
in the range \( 0\leq \sigma \leq \sigma _{c} \). Clearly, \( U^{*}(\sigma _{c})=1/2 \)
(putting \( \sigma =\sigma _{c} \) in Eqn. (1.13)). Therefore the stable
solution can be written as \begin{equation}
\label{may28}
U^{*}(\sigma )=U^{*}(\sigma _{c})+(\sigma _{c}-\sigma )^{1/2};\sigma _{c}=\frac{1}{4}.
\end{equation}
For \( \sigma >\sigma _{c} \) we can not get a real-valued fixed
point as the dynamics never stops until \( U_{t}=0 \), when the bundle
breaks completely. It may be noted that the quantity \( U^{*}(\sigma )-U^{*}(\sigma _{c}) \)
behaves like an order parameter that determines a transition from
a state of partial failure (\( \sigma \leq \sigma _{c} \)) to a state
of total failure (\( \sigma >\sigma _{c} \)) \cite{SBP02}:
\begin{equation}
\label{dec31}
O\equiv U^{*}(\sigma )-U^{*}(\sigma _{c})=(\sigma _{c}-\sigma )^{\beta };\beta =\frac{1}{2}.
\end{equation}

 To study the dynamics away from criticality (\( \sigma \rightarrow \sigma _{c} \)
from below), we replace the recursion relation (1.11) by a differential
equation \begin{equation}
\label{qwes}
-\frac{dU}{dt}=\frac{U^{2}-U+\sigma }{U}.
\end{equation}

\noindent Close to the fixed point we write \( U_{t}(\sigma )=U^{*}(\sigma ) \)
+ \( \epsilon  \) (where \( \epsilon \rightarrow 0 \)). This, following
Eq. (1.13), gives \cite{SBP02} \begin{equation}
\label{qas}
\epsilon =U_{t}(\sigma )-U^{*}(\sigma )\approx \exp (-t/\tau ),
\end{equation}

\noindent where \( \tau =\frac{1}{2}\left[ \frac{1}{2}(\sigma _{c}-\sigma )^{-1/2}+1\right]  \).
Near the critical point we can write \begin{equation}
\label{dec19}
\tau \propto (\sigma _{c}-\sigma )^{-\alpha };\alpha =\frac{1}{2}.
\end{equation}
 Therefore the relaxation time diverges following a power-law as \( \sigma \rightarrow \sigma _{c} \)
from below \cite{SBP02}. 

One can also consider the breakdown susceptibility \( \chi  \), defined
as the change of \( U^{*}(\sigma ) \) due to an infinitesimal increment
of the applied stress \( \sigma  \) \cite{SB01,SBP02} \begin{equation}
\label{sawq}
\chi =\left| \frac{dU^{*}(\sigma )}{d\sigma }\right| =\frac{1}{2}(\sigma _{c}-\sigma )^{-\eta};\eta =\frac{1}{2}
\end{equation}

\noindent from Eqn. (1.13). Hence the susceptibility diverges as
the applied stress \( \sigma  \) approaches the critical value \( \sigma _{c}=\frac{1}{4} \).
Such a divergence in \( \chi  \) had already been observed in the
numerical studies \cite{SB01}. 




\subsubsection{For linearly increasing distribution of fiber strength}
\noindent Here, the cumulative probability becomes \begin{equation}
\label{may20-2}
P(\sigma _{t})=\int ^{\sigma _{t}}_{0}\rho (\sigma _{th})d\sigma _{th}=2\int ^{\sigma _{t}}_{0}\sigma _{th}d\sigma _{th}=\sigma ^{2}_{t}.
\end{equation}
Therefore \( U_{t} \) follows a recursion relation (following Eq.
(1.6)) \begin{equation}
\label{qrw}
U_{t+1}=1-\left( \frac{\sigma }{U_{t}}\right) ^{2}.
\end{equation}

\noindent At the fixed point (\( U_{t+1}=U_{t}=U^{*} \)), the above
recursion relation can be represented by a cubic equation of \( U^{*} \)
\begin{equation}
\label{may20-5}
(U^{*})^{3}-(U^{*})^{2}+\sigma ^{2}=0.
\end{equation}

Solving the above equation we get \cite{SBP02} the value of critical stress 
\( \sigma _{c}=\sqrt{4/27} \) which is the strength of the bundle for the above 
fiber strength distribution. Here, the order parameter, susceptibility, 
relaxation time all follow the same power laws (Eqns. (1.15), (1.18) and 
(1.19)) observed for uniform  strength distribution.


\subsubsection{For linearly decreasing distribution of fiber strength}
\noindent In this case, the cumulative probability becomes \begin{equation}
\label{may20-3}
P(\sigma _{t})=\int ^{\sigma _{t}}_{0}\rho (\sigma _{th})d\sigma _{th}=2\int ^{\sigma _{t}}_{0}(1-\sigma _{th})d\sigma _{th}=2\sigma _{t}-\sigma ^{2}_{t}
\end{equation}
and \( U_{t} \) follows a recursion relation (following Eqn. (1.6))
\begin{equation}
\label{qrw}
U_{t+1}=1-2\frac{\sigma }{U_{t}}+\left( \frac{\sigma }{U_{t}}\right) ^{2}.
\end{equation}

\noindent At the fixed point (\( U_{t+1}=U_{t}=U^{*} \)), the above
recursion relation can be represented by a cubic equation of \( U^{*} \)
\begin{equation}
\label{may20-5}
(U^{*})^{3}-(U^{*})^{2}+2\sigma U^{*}-\sigma ^{2}=0.
\end{equation}

Solution of the above equation gives \cite{SBP02} the value of critical stress
\( \sigma _{c}=4/27 \) which is the strength of the
bundle. Here also, the precursor parameters, the order parameter,
 susceptibility and  relaxation time,  all follow the same power laws 
(Eqns. (1.15), (1.18) and (1.19)) observed for uniform strength distribution.


Thus the democratic fiber bundles (for different fiber strength distributions)
show phase transition from a state of partial failure to total failure
with a well defined precursors (order parameter, susceptibility and
relaxation time) which show similar power law variation on the way as
the critical point is approached; characterized by the universal values
of the associated exponents (\( \alpha  \), \( \beta  \)and  \( \eta  \)).
All the above scaling behaviors represented by Eqns. 
(1.15), (1.18) and (1.19) for stress (\( \sigma \)) below the global failure 
stress (\( \sigma_{c} \)) 
of the bundle suggests that a prior 
knowledge of the responses like the fraction of failed fibers, relaxation 
time etc. can be extrapolated following the above power laws to estimate the 
global failure point \( \sigma_{c} \). 

\vskip .2in

\section{Two-fractal overlap model of earthquake and
the prediction possibility of large events}

\noindent The two-fractal overlap model of earthquake  \cite{BS99} is a
very recent modeling attempt.
Such  models are all  based on the observed `plate tectonics'
and the fractal nature of the interface between tectonic plates
and earth's solid crust.
The statistics of overlaps between two fractals is
not studied much yet, though their knowledge is often required in various
physical contexts. For example, it has been established recently that since
the fractured surfaces have got well-characterized self-affine properties
\cite{BB82}, the distribution of the elastic energies
released during the slips between two fractal surfaces (earthquake
events) may follow the overlap distribution of two self-similar fractal
surfaces \cite{BS99,SBP03}. Chakrabarti and
Stinchcombe \cite{BS99}
had shown analytically that for regular fractal overlap (Cantor sets
and carpets) the contact area distribution follows a simple power
law decay.
The two fractal overlap magnitude changes in time as one fractal moves
over the other. The overlap (magnitude)
time series can therefore be studied as a model time series of earthquake
avalanche dynamics \cite{CL89}.

Here, we study the time (\( t \)) variation of contact area (overlap)
 \( m(t) \) between two well-characterized fractals having the same
fractal dimension as one fractal moves over the other with constant
velocity. We have chosen only very simple fractals: regular and random
Cantor sets, gaskets and percolating cluster. We analyse the time
series data of Cantor set overlaps only to find the prediction possibility
of large events (occurrence of large overlaps). We show that the time
series \( m(t) \) obtained by moving one fractal uniformly over the
other (with periodic boundary condition) has some features which can be
utilized to predict the
`large events'.  This is shown utilizing
the discrete or a `quantum' nature of the integrated (cumulative)
overlap over time.

\vskip.2in
\subsection{Two-fractal overlaps:}
\noindent We now discuss the two-fractal overlap sizes. Let us take two 
identical Cantor sets at finite generation and slide one over the other, 
 assuming periodic boundary condition. Two such cases for a regular Cantor 
sets as well as a gasket are shown in Fig. 5. Similar cases for random 
fractals are shown in Fig. 6.

\resizebox*{4.5cm}{3cm}{\includegraphics{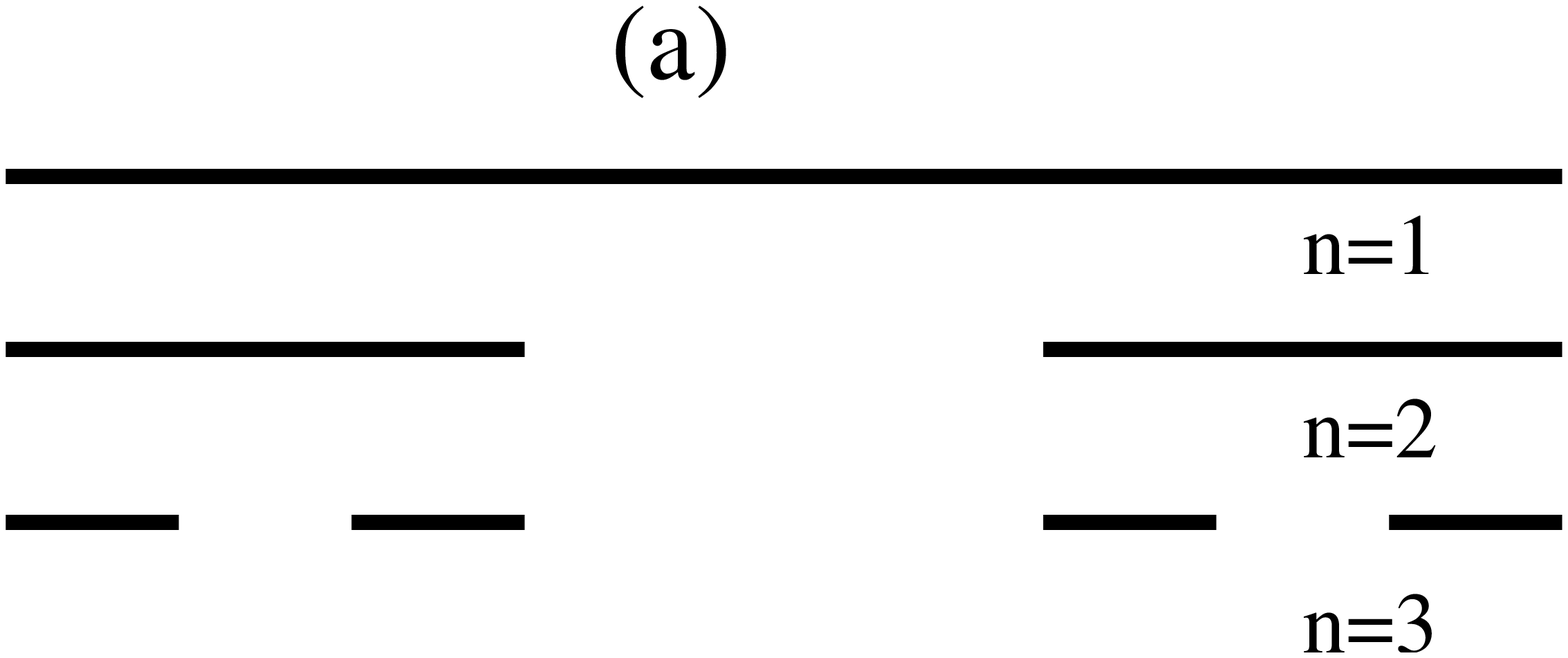}} \hskip.3in
\resizebox*{4.5cm}{3cm}{\includegraphics{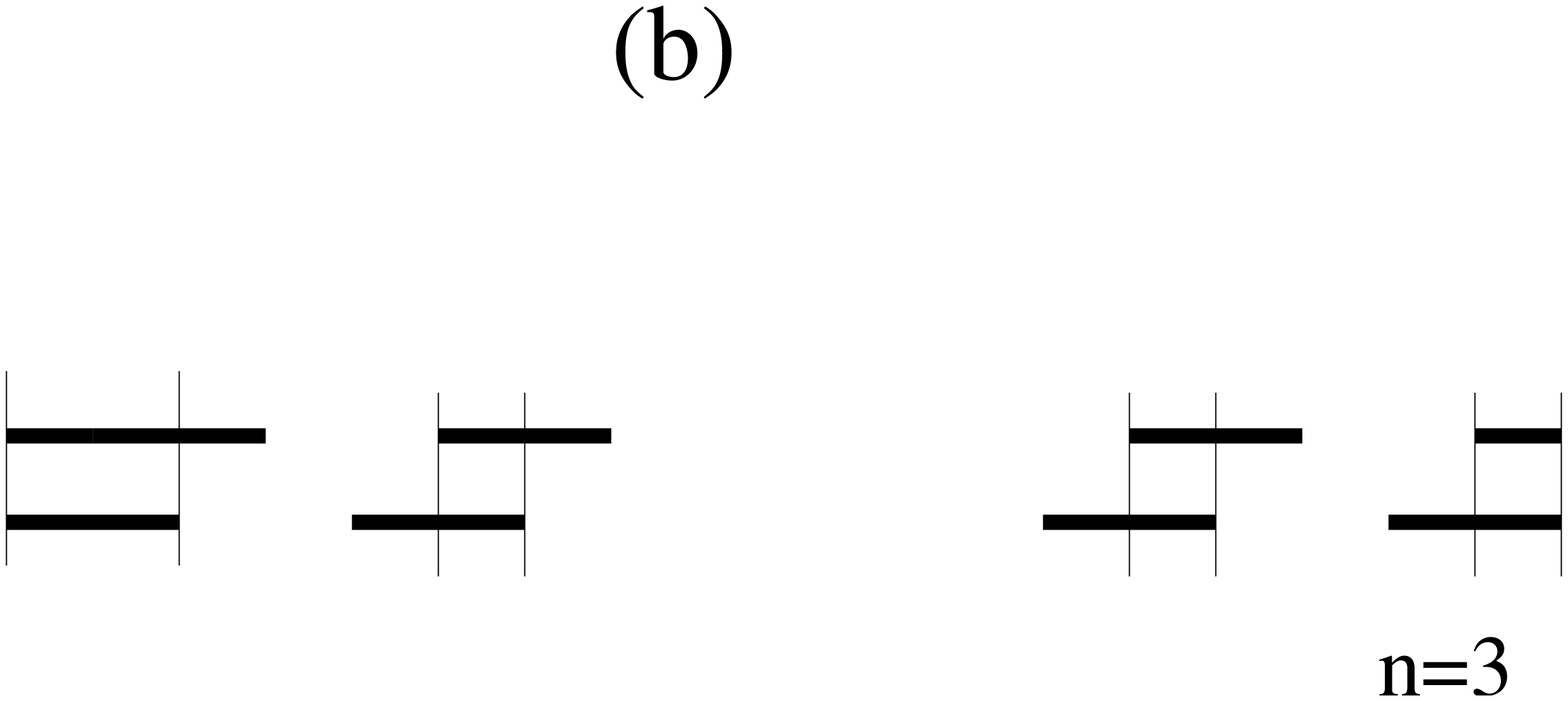}} 

\vskip.1in

\resizebox*{4cm}{3.5cm}{\rotatebox{-90}{\includegraphics{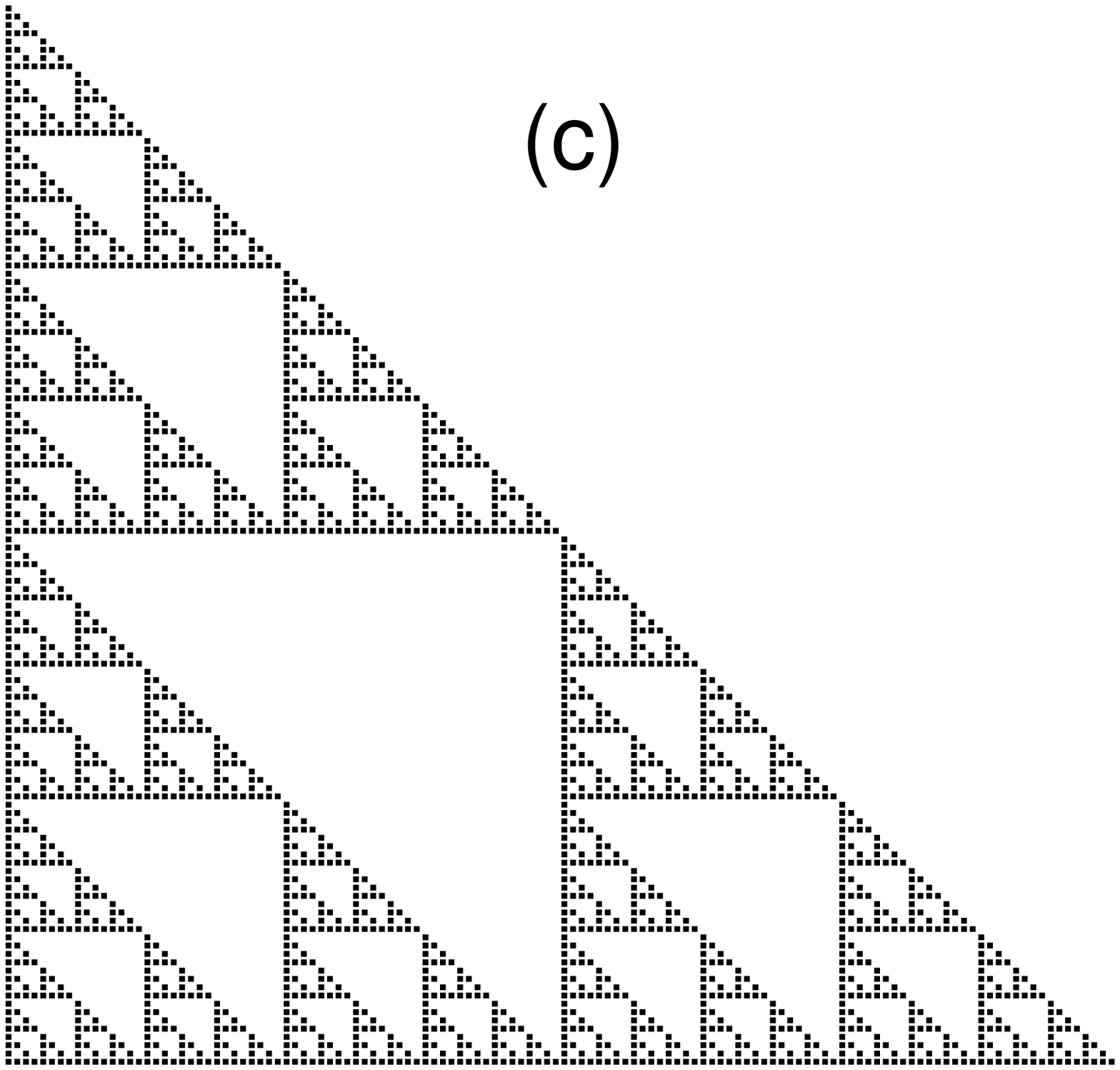}}} 
\hskip.6in\resizebox*{4cm}{3.5cm}{\rotatebox{-90}{\includegraphics{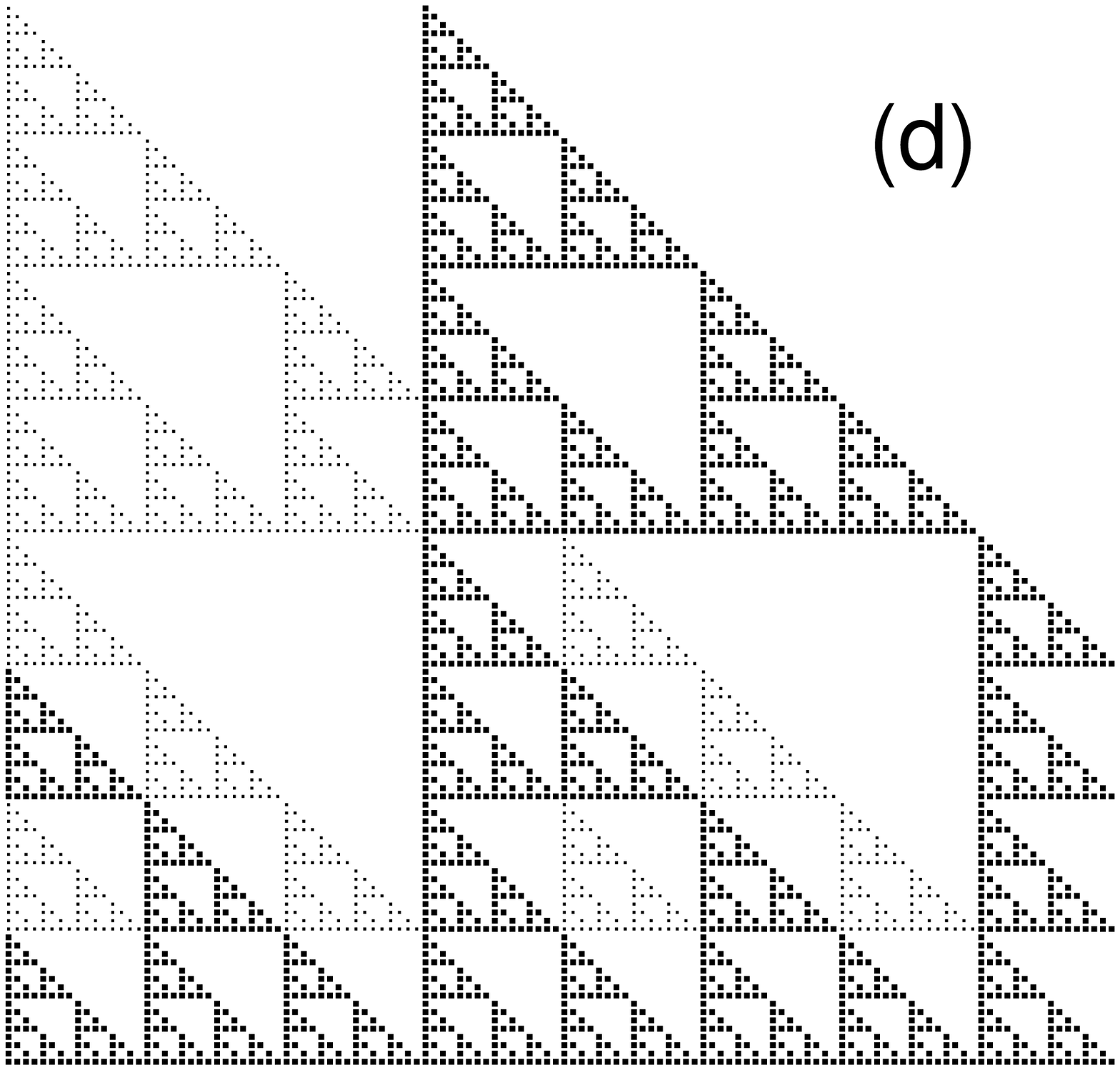}}} 

\vskip.1in

\noindent\textbf{\footnotesize \it Figure 5.} {\footnotesize (a) A regular 
Cantor set of dimension \( \ln 2/\ln 3 \); only three finite generations
are shown. (b) The overlap of two identical (regular) Cantor sets,
at \( n=3 \), when one slips over other; the overlap sets are indicated
within the vertical lines, where periodic boundary condition has been
used. (c) A regular gasket of dimension \( \ln 3/\ln 2 \) at the
\( 7 \)th generation. (d) The overlap of two identical regular gaskets
at same generations (\( n=7 \)) is shown as one is translated over
the other; periodic boundary condition has been used for the translated
gasket. }{\footnotesize \par}


\resizebox*{5cm}{2cm}{\includegraphics{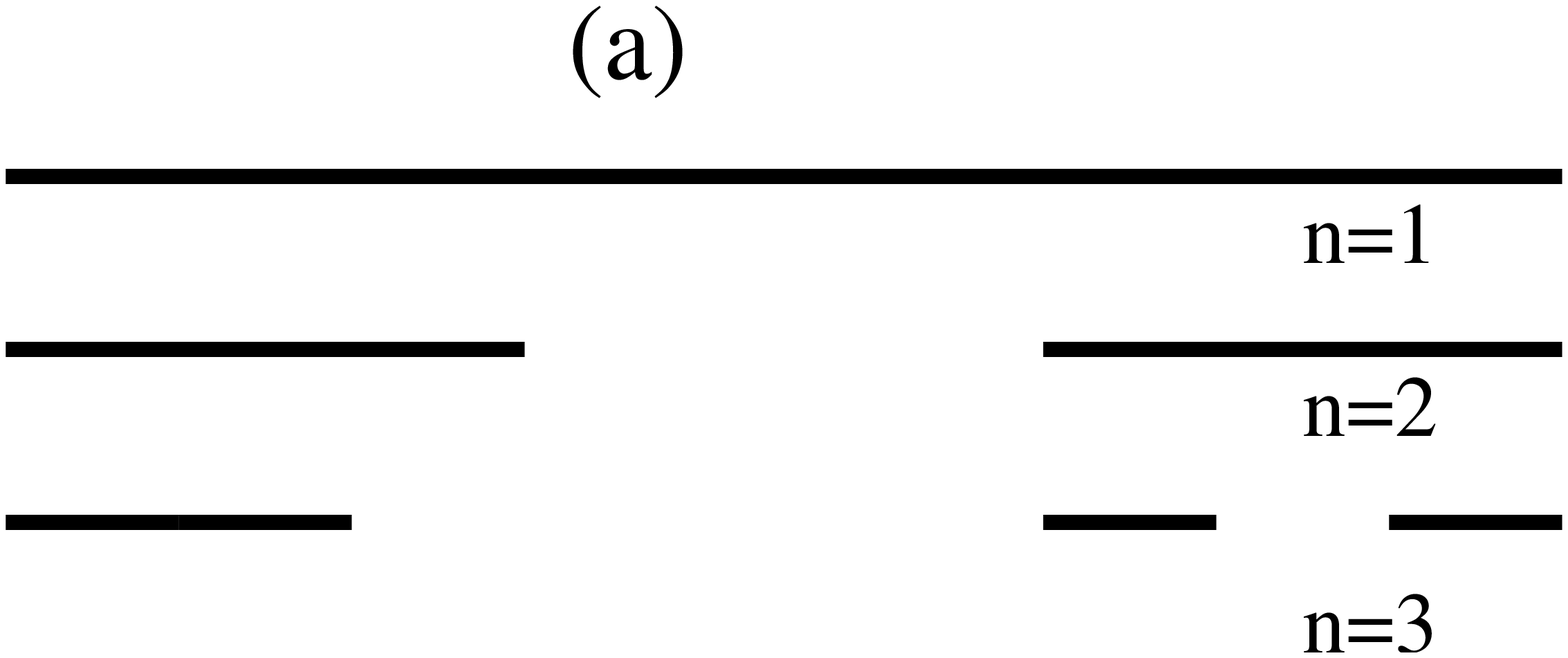}} \hskip.3in
\resizebox*{5cm}{2cm}{\includegraphics{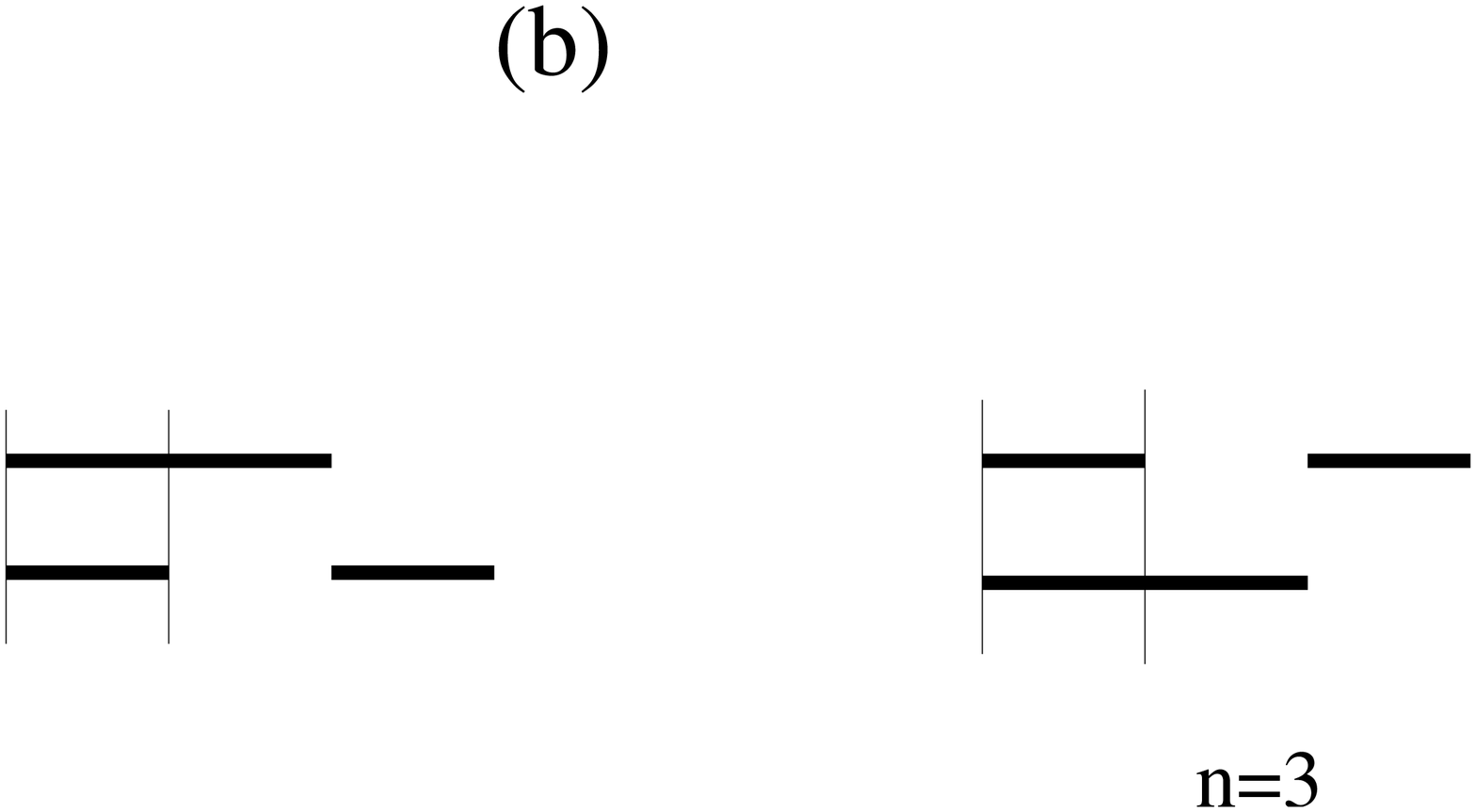}} 

\vskip.1in

\resizebox*{4cm}{3.5cm}{\rotatebox{-90}{\includegraphics{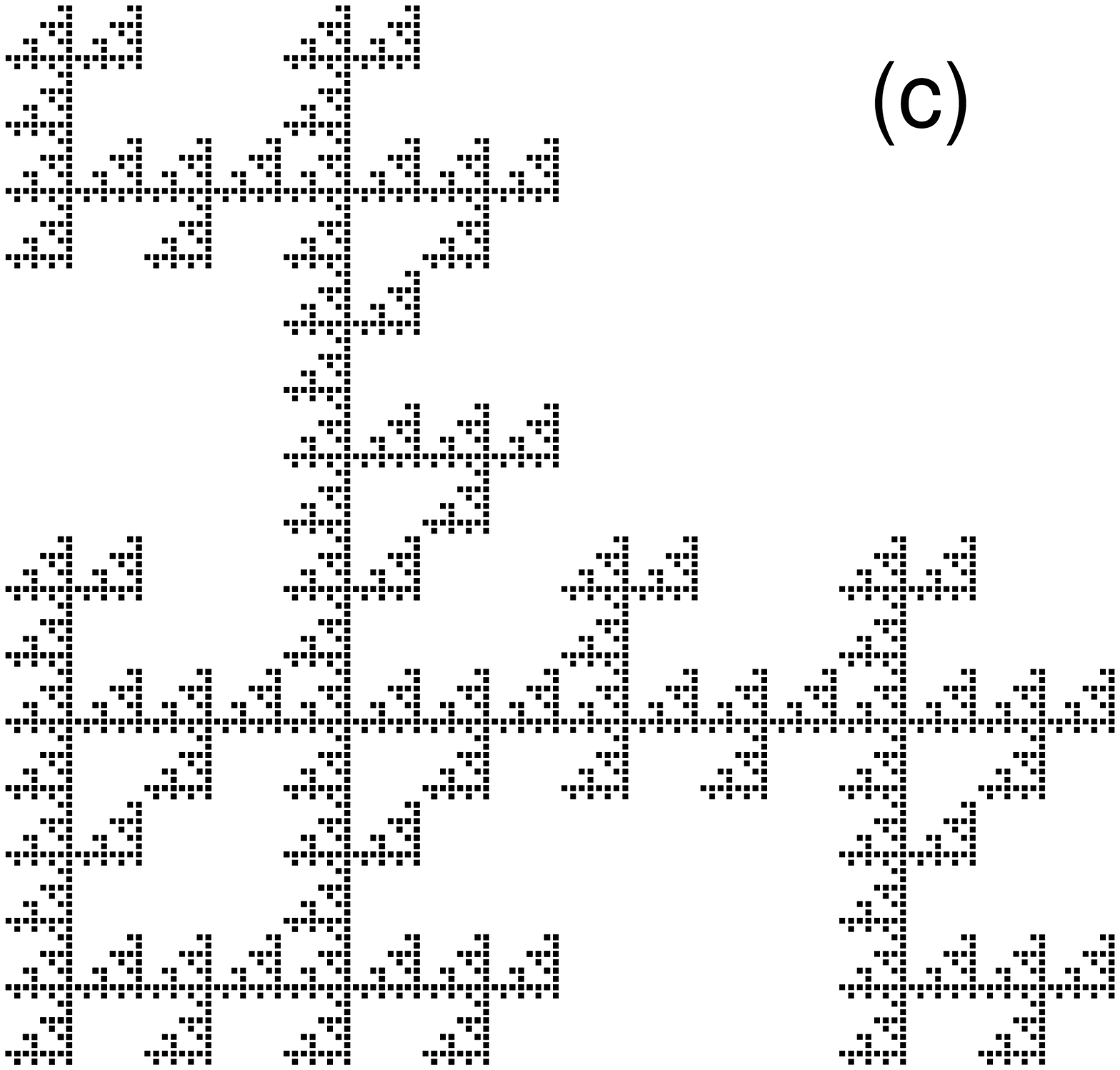}}} 
\hskip.6in\resizebox*{4cm}{3.5cm}{\rotatebox{-90}{\includegraphics{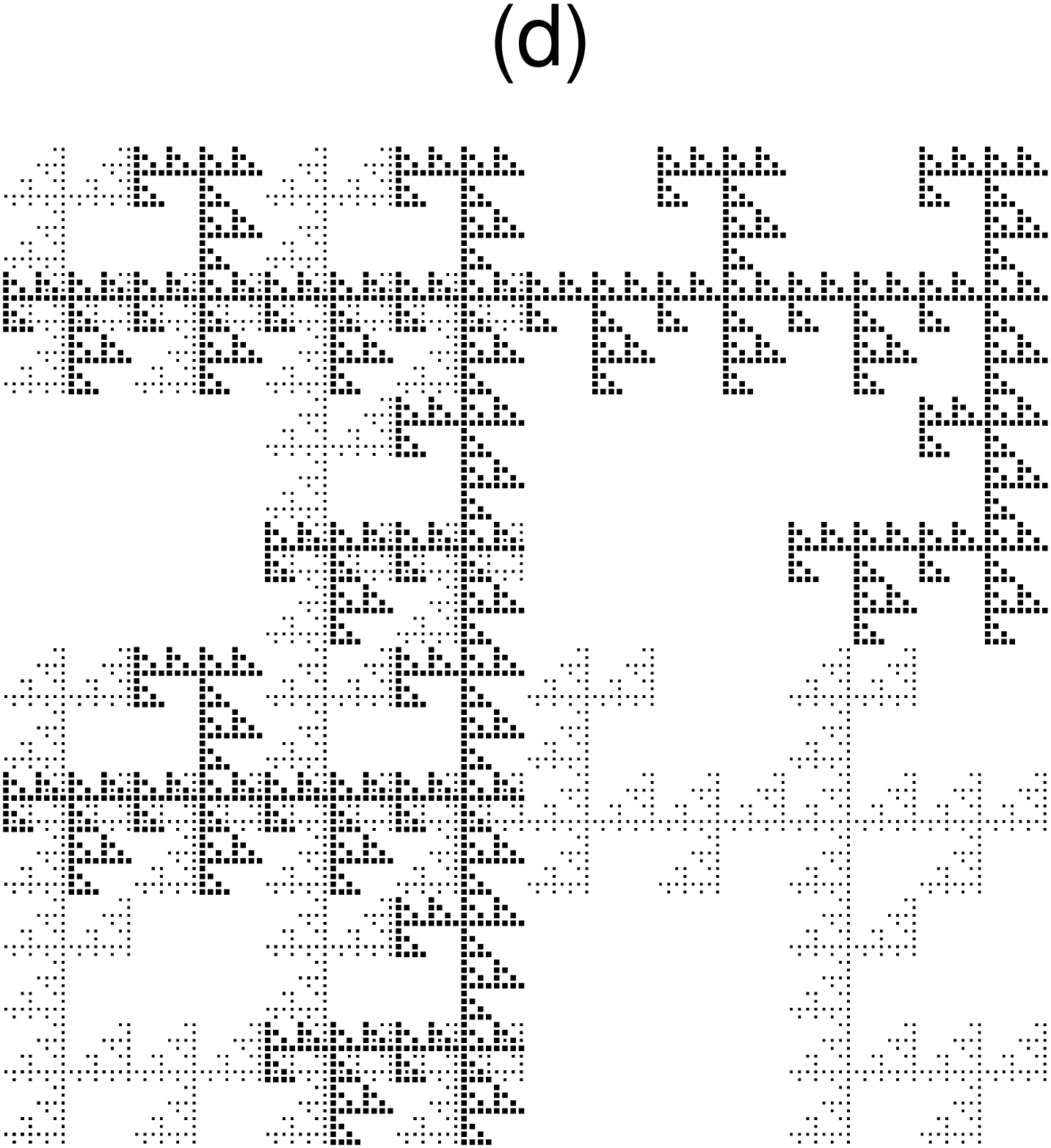}}} 

\vskip.1in

\noindent\textbf{\footnotesize \it Figure 6.} {\footnotesize (a) A random 
Cantor set of dimension \( \ln 2/\ln 3 \); only three finite generations
are shown. (b) Overlap of two random Cantor sets (at \( n=3 \); having
the same fractal dimension) in two different realisations. The overlap
sets are indicated within the vertical bars. (c) A random realisation
of a gasket of dimension \( \ln 3/\ln 2 \) at \( 7 \)th generation.
(d) The overlap of two random gaskets of same dimension and of same
generation but generated in different realisations.}{\footnotesize \par}

\vskip.1in


\noindent Next, we study the overlap distribution of two well-characterised
random fractals; namely the percolating fractals \cite{Stauffer92}.
It seems, although many detailed features of the clusters will change 
with the changes in (parent) fractals, the subtle features of the 
overlap distribution function remains unchanged.

\resizebox*{5cm}{5cm}{\rotatebox{-90}{\includegraphics{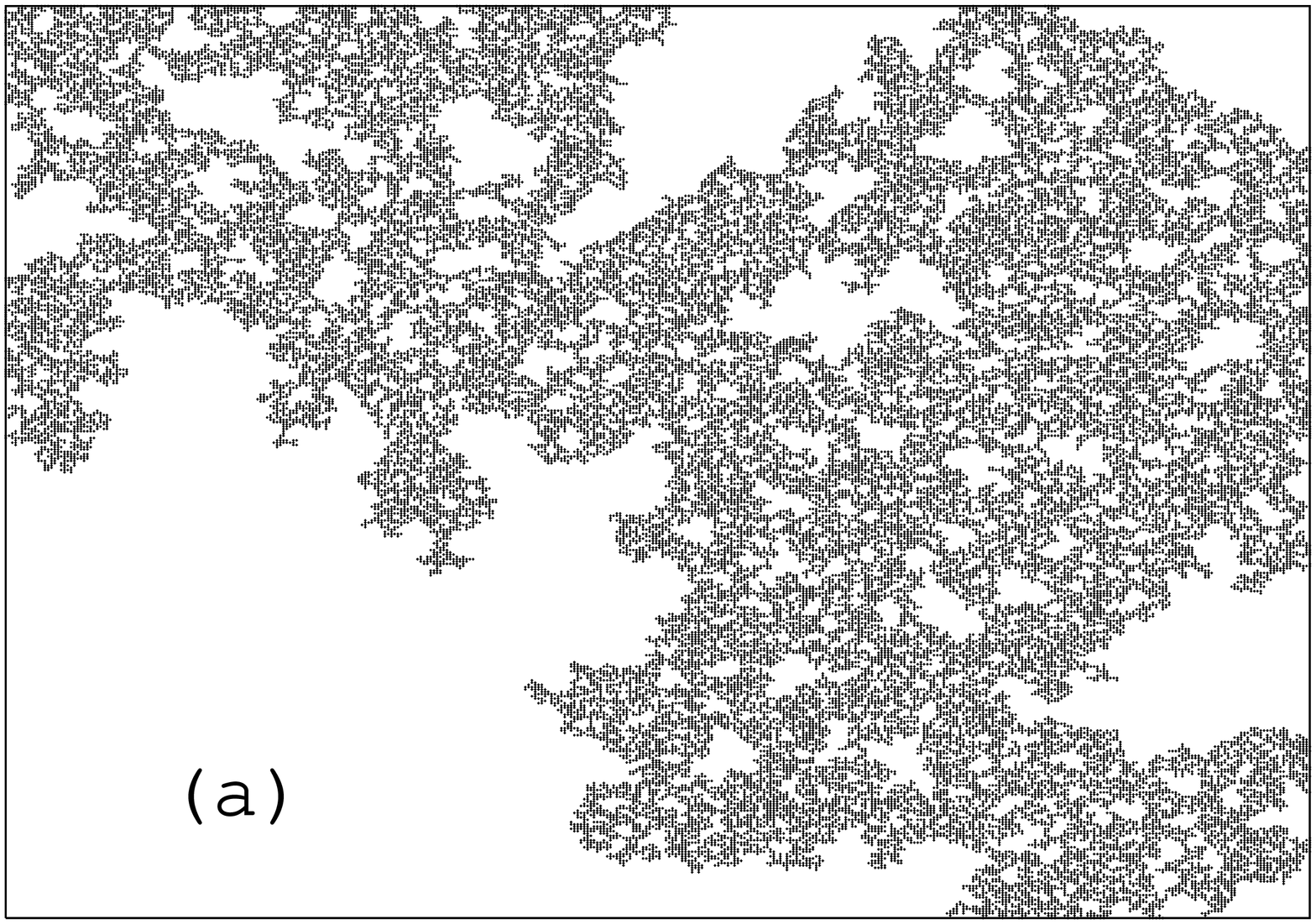}}} 
\hskip.3in\resizebox*{5cm}{5cm}{\rotatebox{-90}{\includegraphics{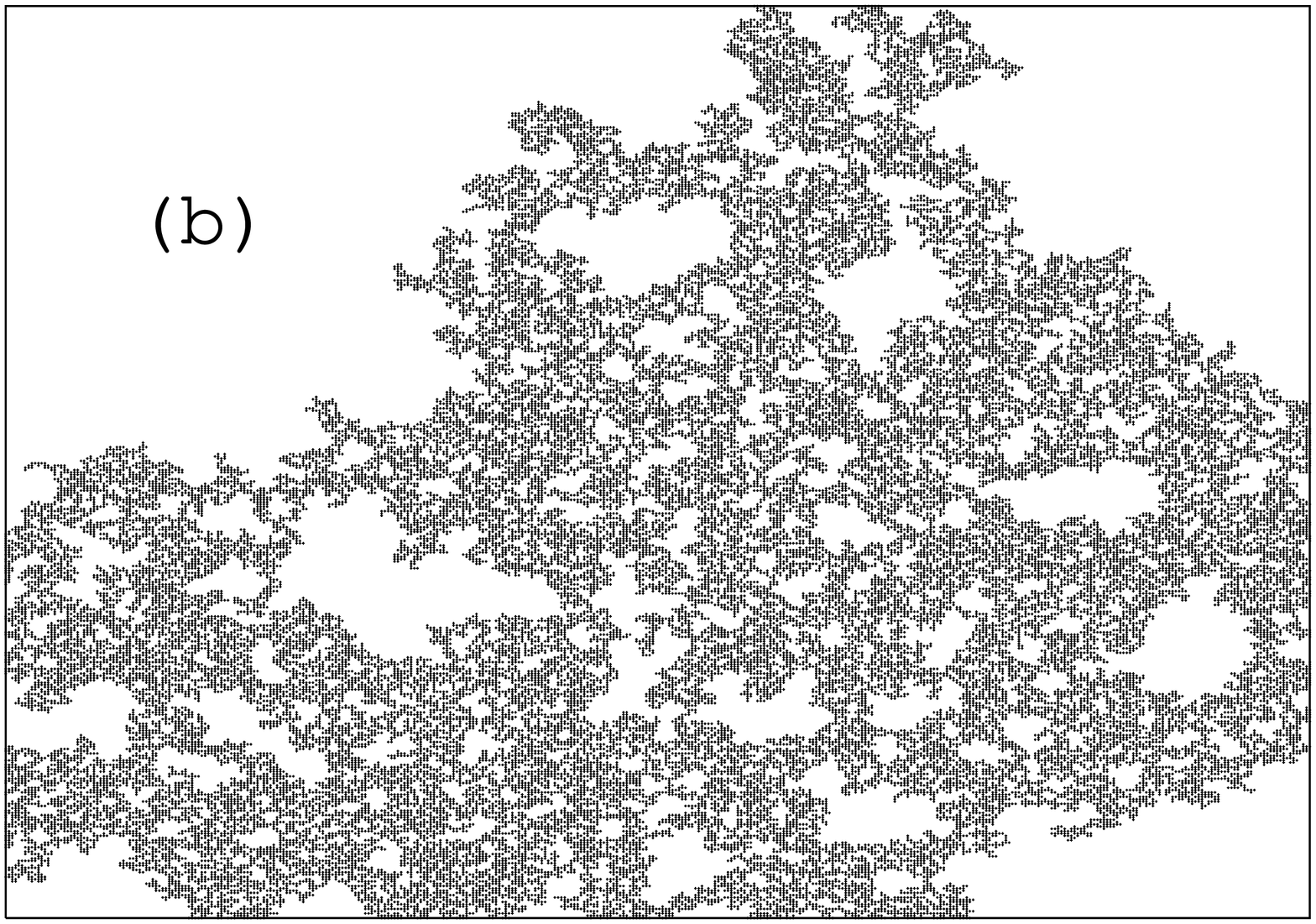}}} 

\vspace{0.1cm}
{\centering \resizebox*{5cm}{5cm}{\rotatebox{-90}{\includegraphics{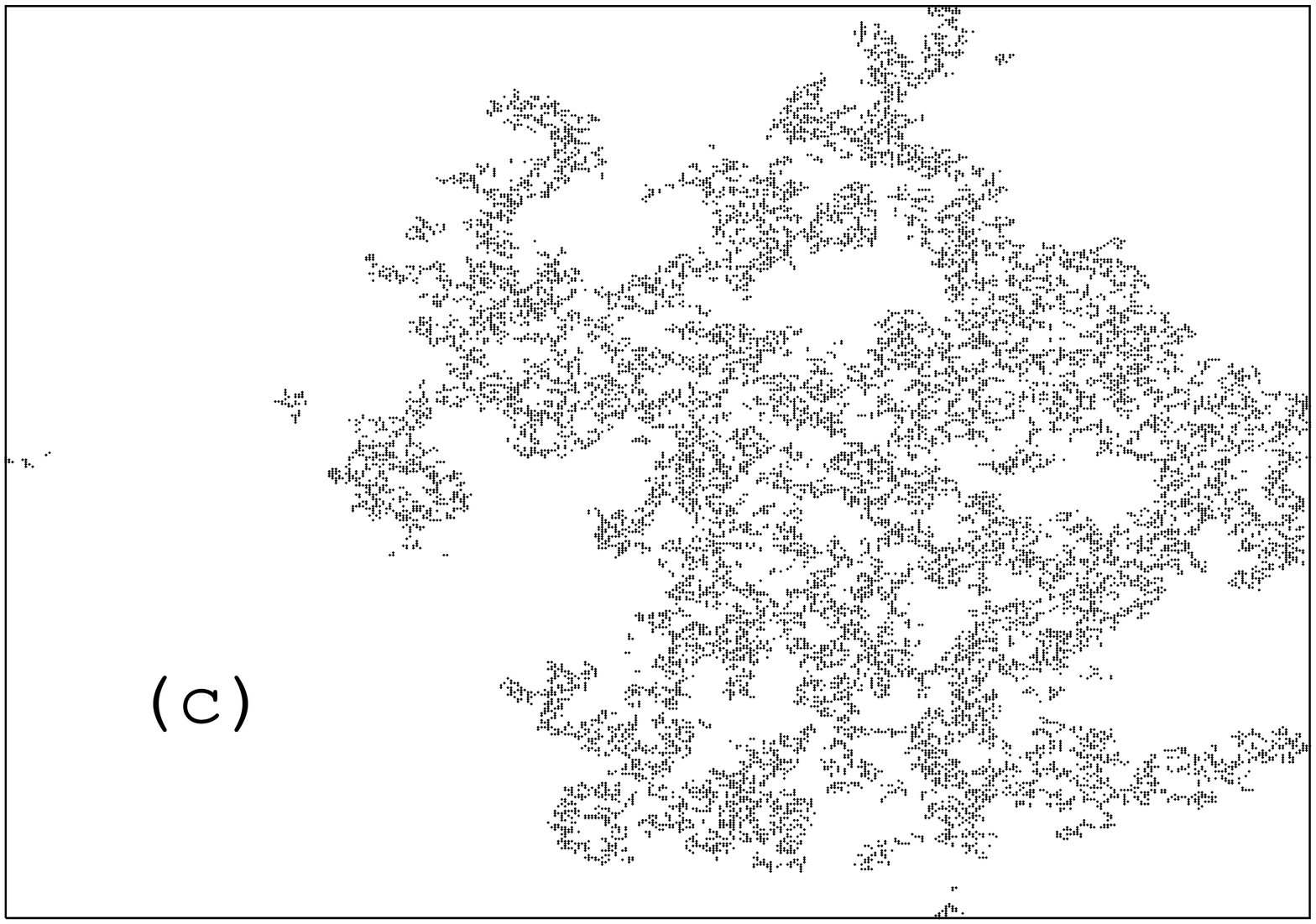}}} \par}

\vskip.1in

\noindent\textbf{\footnotesize \it Figure 7.} {\footnotesize The overlap 
between two percolating clusters; (a) and (b) are two typical realisations
of the same percolating fractal on square lattice (\( d_{f}\simeq 1.89 \))
and (c) their overlap set. Note, the overlap set need not be a connected
one. }{\footnotesize \par}

\vskip.1in

Our earlier study \cite {SBP02} on the overlap statistics for regular and random
Cantor sets, gaskets and percolating clusters indicated 
a universal scaling behavior of the overlap
or contact area (\( m \)) distributions \( P(m) \) for all types of fractal set
overlaps mentioned:  $P^{\prime }(m^{\prime })=L^{\alpha }P(m,L);
m^{\prime }=mL^{-\alpha },$
where \( L \) denotes the finite size of the fractal and the exponent
\( \alpha =2(d_{f}-d) \); \( d_{f} \) being the mass dimension of
the fractal and \( d \) is the embedding dimension. Also the overlap
distribution \( P(m) \), and hence the scaled distribution \( P^{\prime }(m^{\prime }) \),
are seen to decay with \( m \) or \( m^{\prime } \) following a power
law (with exponent value equal to the embedding dimension of the fractals)
for both regular and random Cantor sets and gaskets:
$P(m)\sim m^{-\beta };\beta =d.$
However, for the percolating clusters \cite{Stauffer92}, the overlap size
distribution takes a Gaussian form.


 We consider now the time series obtained by counting the
overlaps \( m(t) \) as a function of time as one Cantor set moves
over the other (periodic boundary condition is assumed) with uniform velocity.
The analysis presented here follows our previous study \cite{Isr-03}.

\subsubsection{Overlap time series data}

\noindent The time series are shown in Fig. 8, for finite generations
of Cantor sets of dimensions \( \ln 2/\ln 3 \) and \( \ln 4/\ln 5 \)
respectively.

\resizebox*{5.25cm}{5cm}{\includegraphics{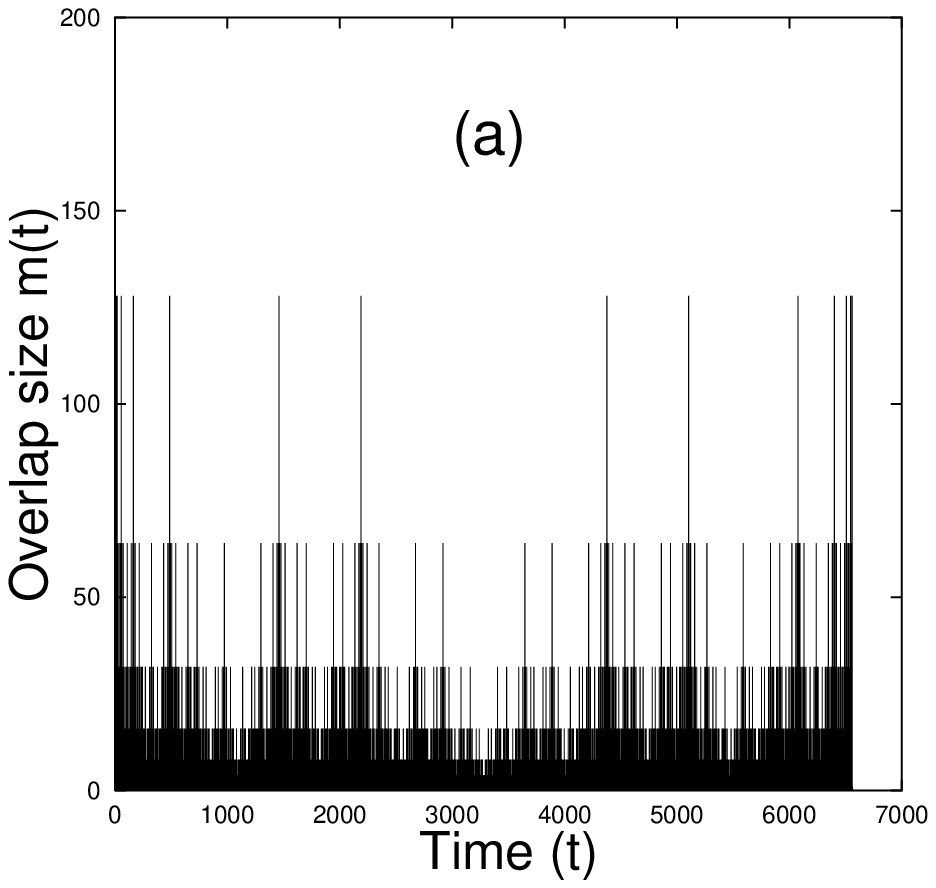}} \hskip.2in
\resizebox*{5.25cm}{5cm}{\includegraphics{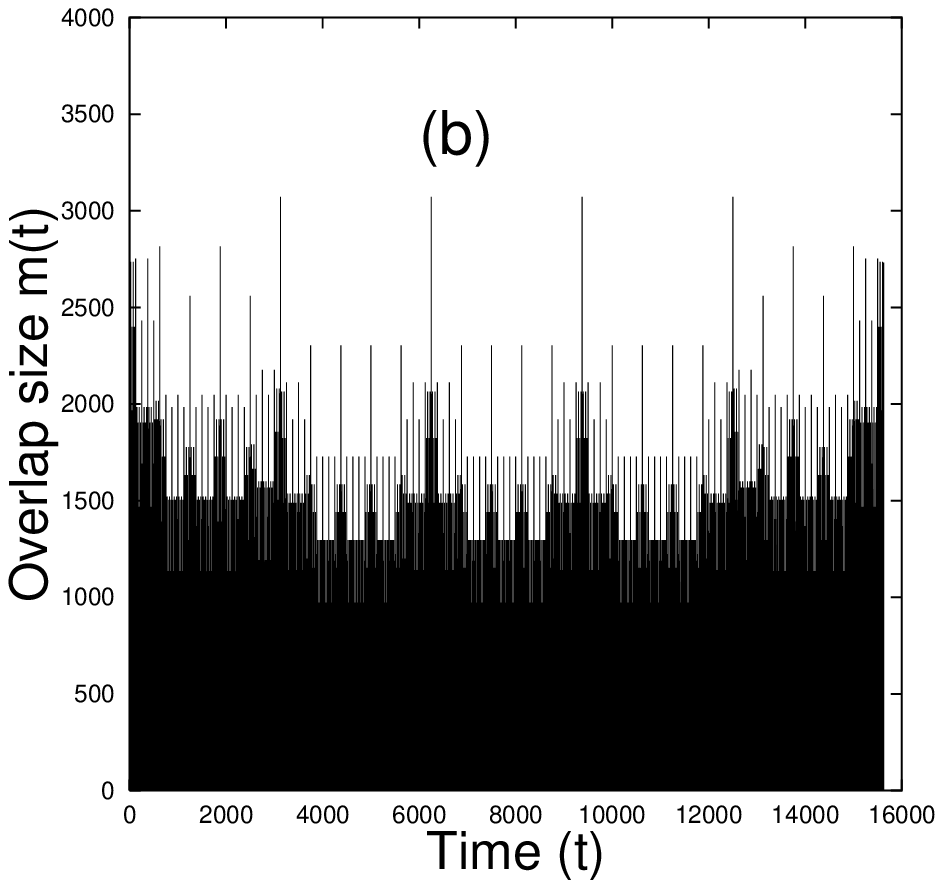}}

\vskip.1in
\noindent\textbf{\small \it Figure 8.} {\small The time (\( t \)) series data ofoverlap size (\( m \)) for regular Cantor sets: (a) of dimension
\( \ln 2/\ln 3 \), at \( 8 \)th generation: (b) of dimension \( \ln 4/\ln 5 \),
at \( 6 \)th generation. The obvious periodic repeat of the time series
comes from the assumed periodic boundary condition of one of the
sets (over which the other one slides).}{\small \par}

\subsubsection{Cumulative overlap sizes}

\noindent Now we calculate the cumulative overlap size
$Q(t)=\int _{o}^{t}mdt$
\noindent and plot that against time in Fig. 9. Note, that `on average'
\( Q(t) \) is seen to grow linearly with time \( t \) for regular
as well as random Cantor sets. This gives a clue that instead of looking
at the individual overlaps \( m(t) \) series one may look for the
cumulative quantity. 
We observe \( Q(t)\simeq ct \), where \( c \) is
dependent on the fractal. This result is even more prominent in the case
of Cantor sets with \( d_{f}= \) \( \ln 4/\ln 5 \).

\vskip.2in
\resizebox*{5cm}{4.5cm}{\includegraphics{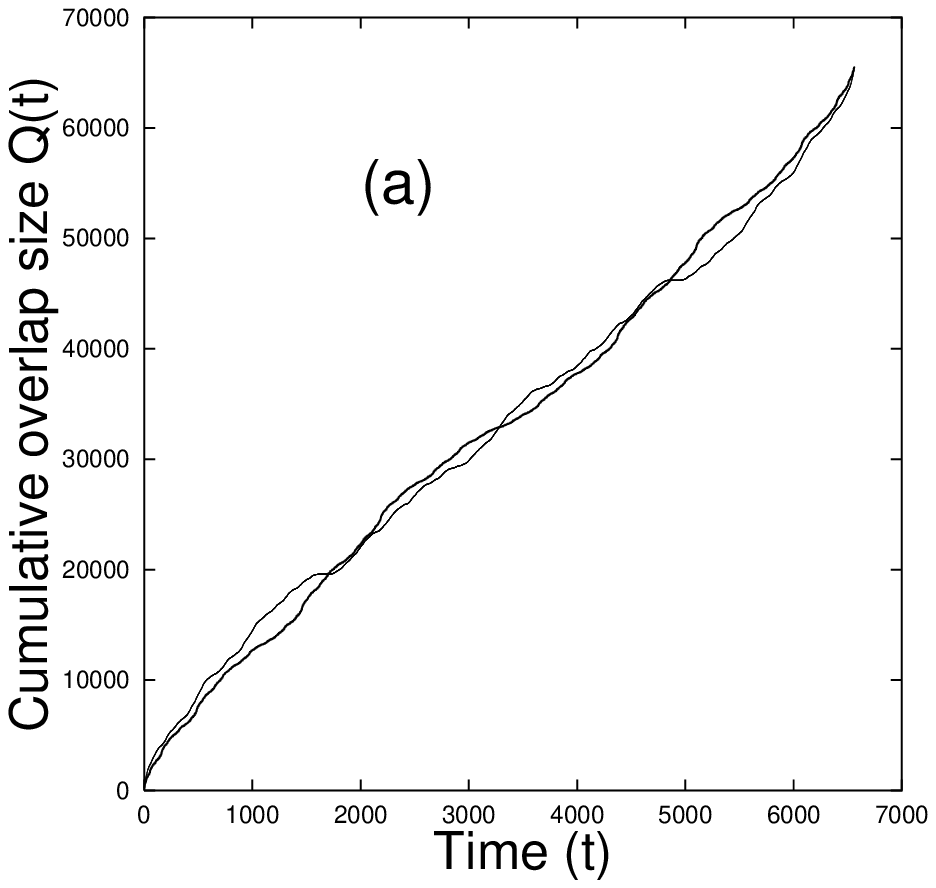}} \hskip.2in\resizebox*{5cm}{4.5cm}{\includegraphics{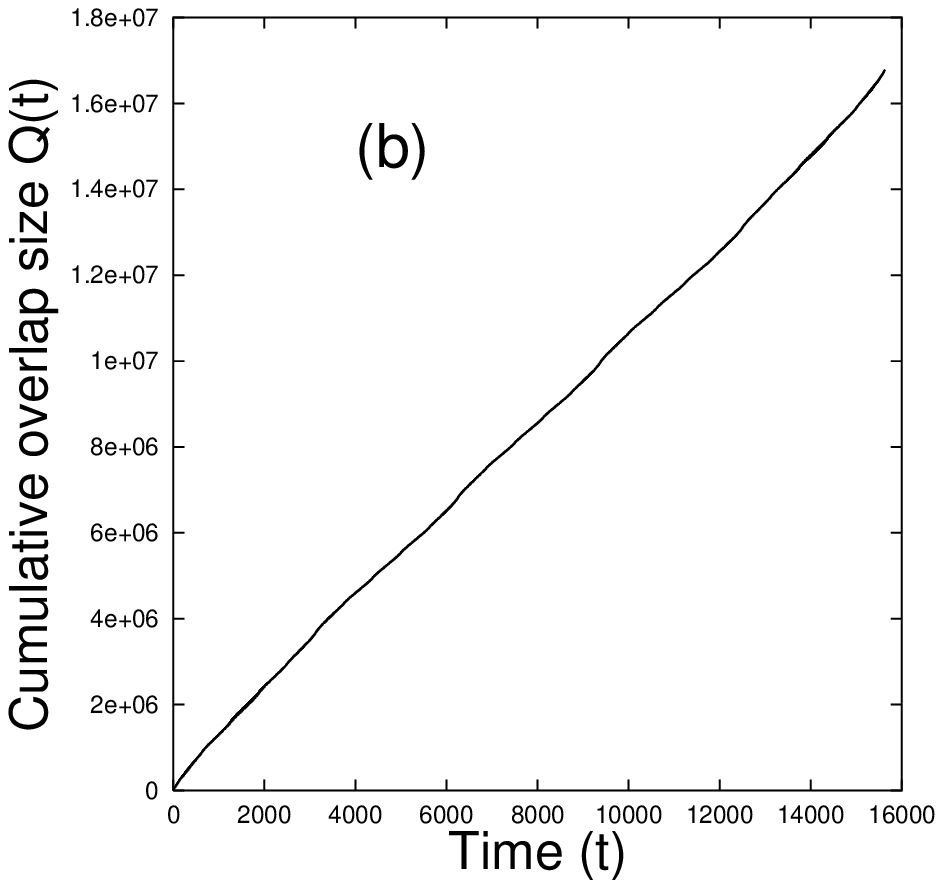}}

\vskip.1in
\noindent\textbf{\small \it Figure 9.} {\small The cumulative overlap \( Q \) versus
time; for pure cantor sets: (a) of dimension \( \ln 2/\ln 3 \) (at
\( 8 \)th generation) and (b) of dimension \( \ln 4/\ln 5 \) (at
\( 6 \)th generation). The dotted line corresponds to those for two identical
but random Cantor sets. In (b) the two lines fall on each other.}{\small \par}

\subsubsection{Cumulative overlap quantization}

\noindent We first identify the `large events' occurring at time \( t_{i} \)
in the \( m(t) \) series, where \( m(t_{i})\geq M \), a pre-assigned
magnitude. We then look for the cumulative overlap size
\( Q_{i}(t)=\int _{t_{i-1}}^{t}mdt \), \( t\) \( \leq t_{i}\),
where the successive large events occur at times \( t_{i-1} \) and
\( t_{i} \). The behavior of \( Q_{i} \) with time is shown in Fig. 10 
 for regular cantor sets with \( d_{f}=  \) \( \ln 2/\ln 3 \) at
generation \( n=8 \). Similar results are also given for Cantor sets
with \( d_{f}=  \) \( \ln 4/\ln 5 \) at generation \( n=6 \) in Fig. 11.
 \( Q_{i} (t)\) is seen to grow almost linearly in time up to \(Q_{i}(t_{i}\))
after which it drops down to zero. It appears that there are discrete
(quantum) values of \( Q_{i} (t_{i}\)). One can therefore anticipate large 
events when the cumulative seismic overlap \( Q_{i} (t)\), since the last 
major event at \( t_{i-1} \), grows linearly with time \( t\) to the 
discrete 
levels \( lQ_{0} \), specific to the series of events. No major event occurs 
during the growth period of \( Q_{i}(t) \) for the intermediate values. 

\resizebox*{5.5cm}{5.5cm}{\includegraphics{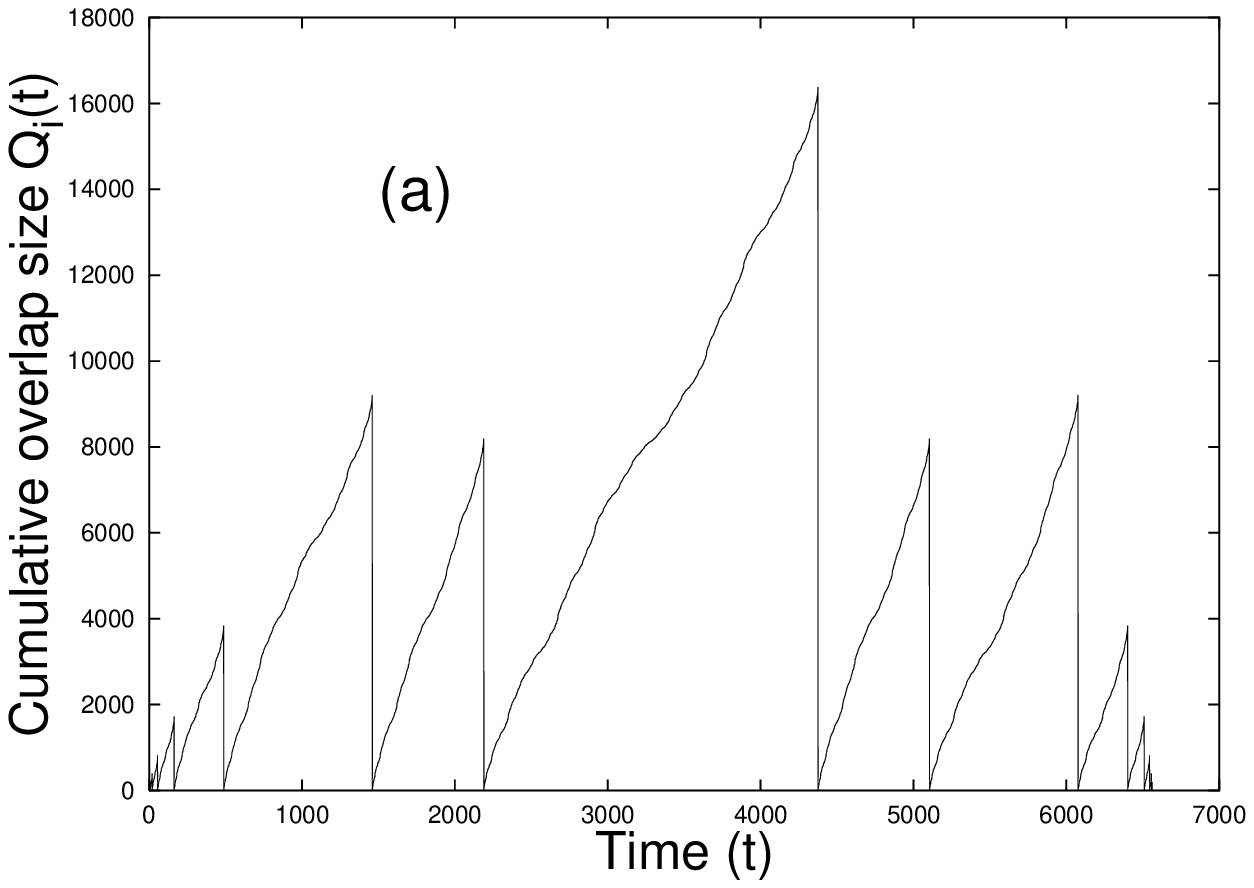}} \hskip.3in\resizebox*{5.5cm}{5.5cm}{\includegraphics{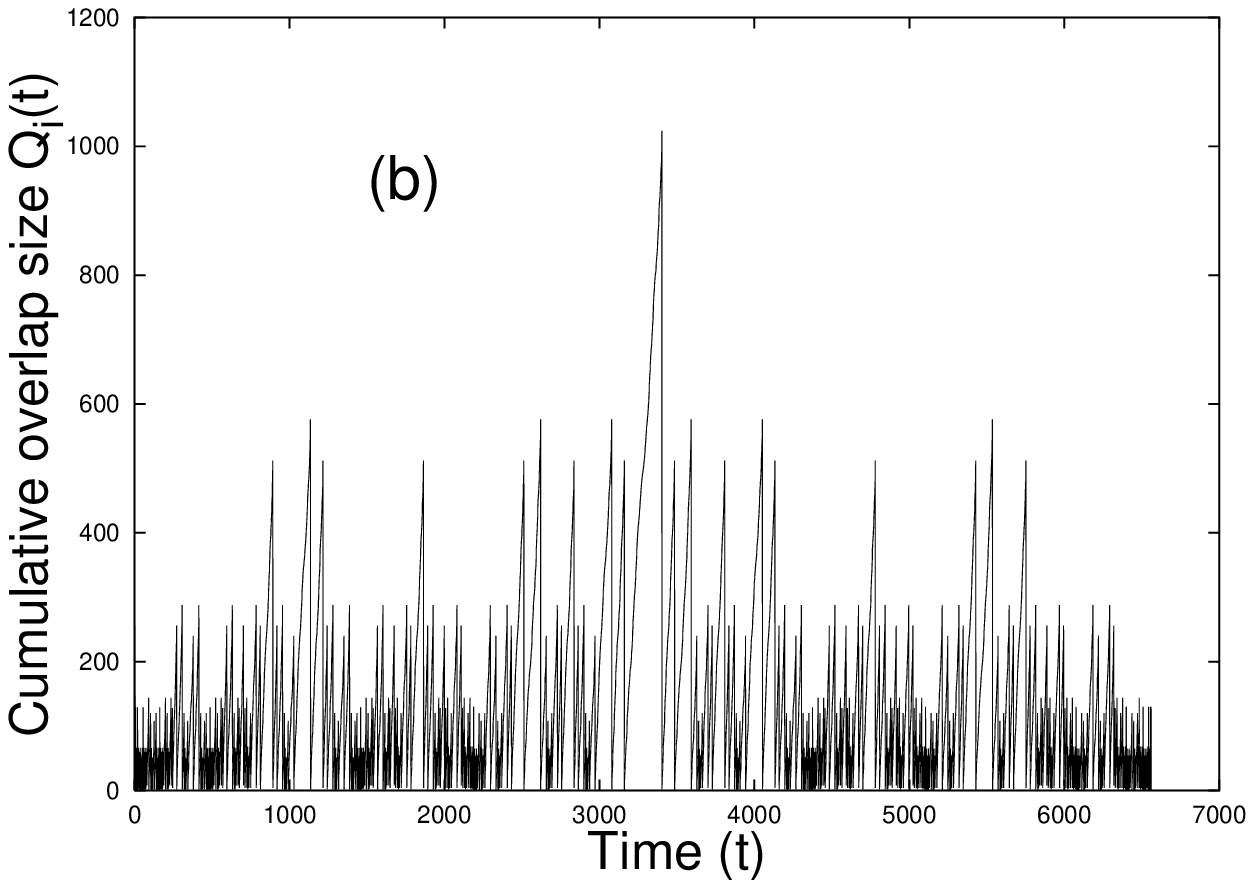}}

\vskip.1in

\noindent\textbf{\small \it Figure 10.} {\small The cumulative overlap
size variation
with time (for regular Cantor sets of dimension \( \ln 2/\ln 3 \),
at \( 8 \)th generation), where the cumulative overlap has been reset
to \( 0 \) value after every big event (of overlap size \( \geq M \)
where \( M=128\) (a) and \( 32 \) (b) respectively).}

\resizebox*{5.5cm}{5.5cm}{\includegraphics{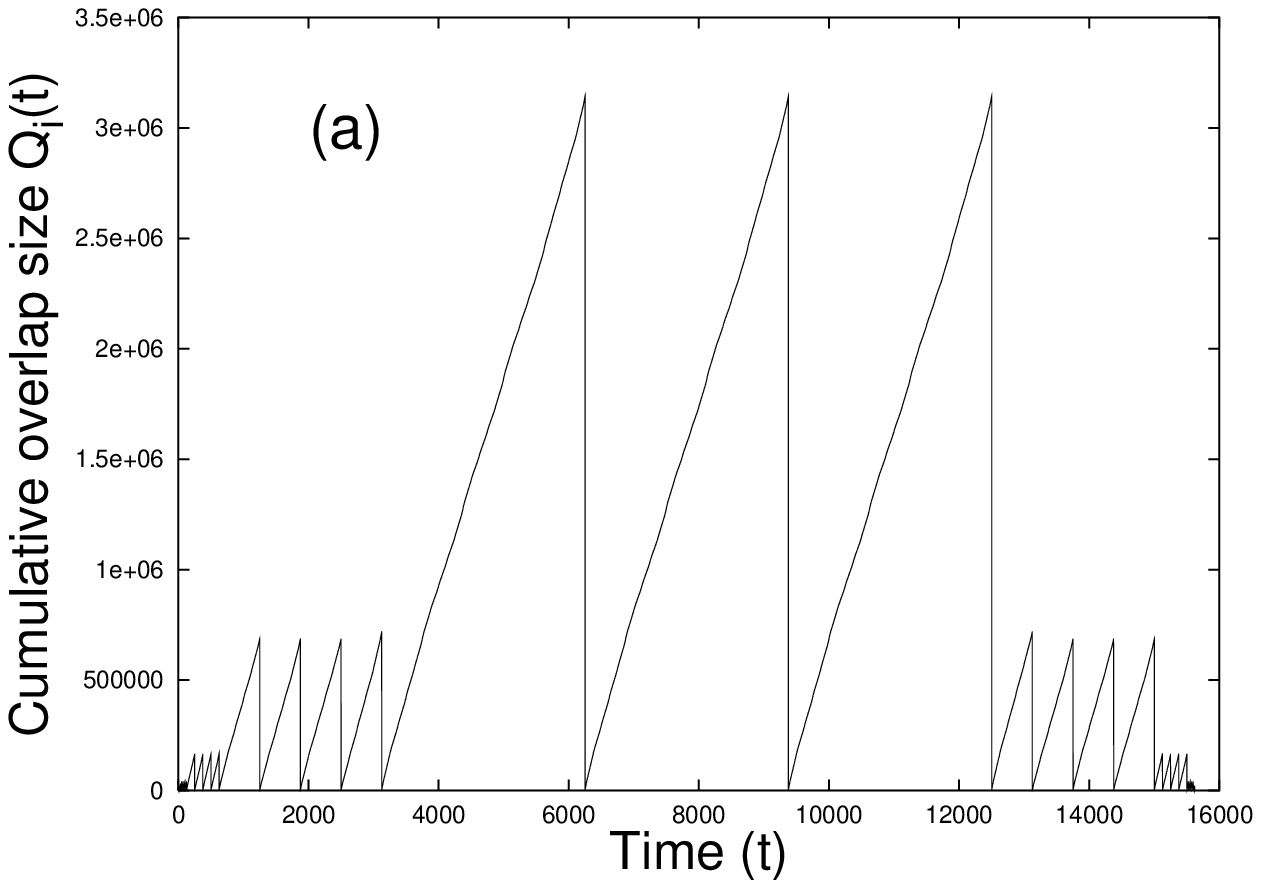}} \hskip.2in\resizebox*{5.5cm}{5.5cm}{\includegraphics{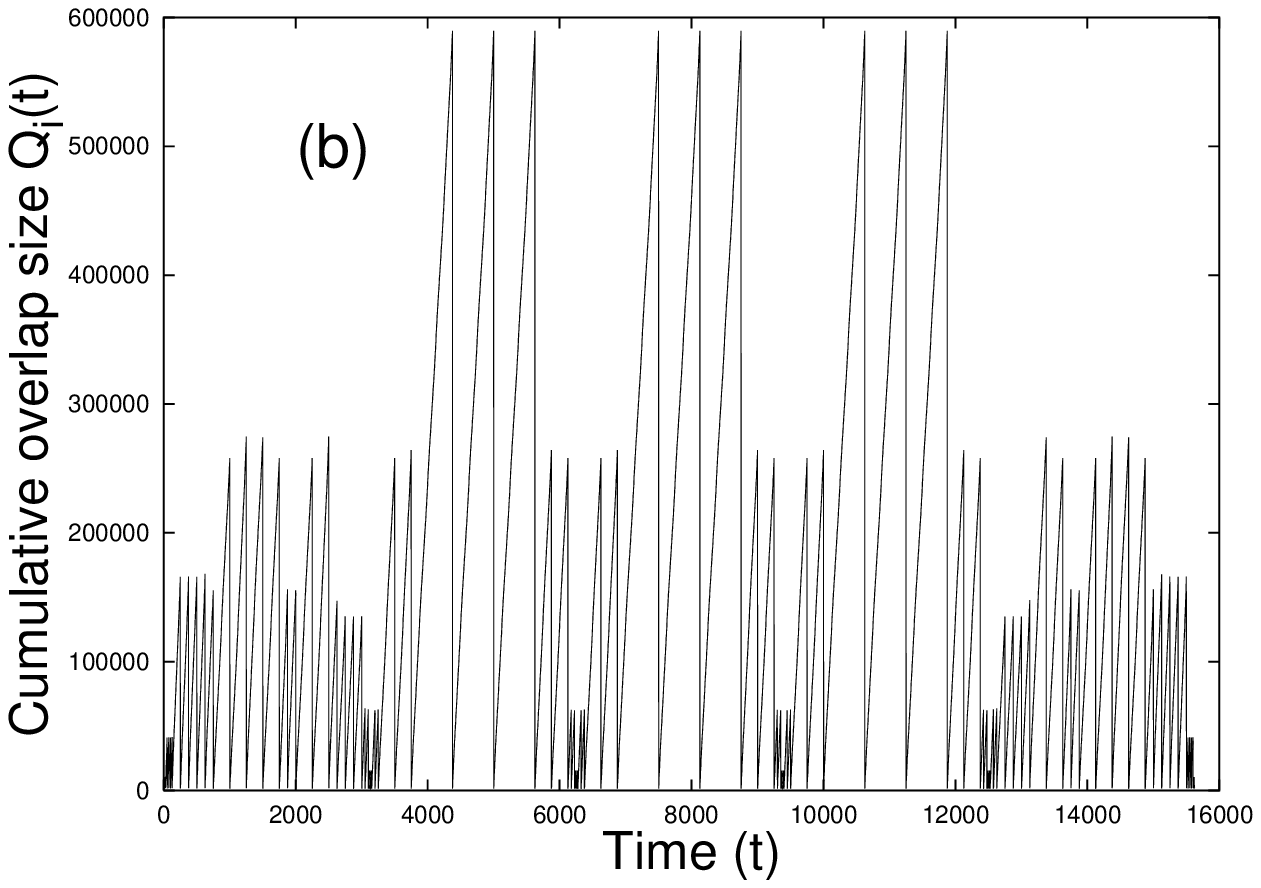}}

\noindent\textbf{\small \it Figure 11.} {\small The cumulative overlap
size variation
with time (for regular Cantor sets of dimension \( \ln 4/\ln 5 \),
at \( 6 \)th generation), where the cumulative overlap has been reset
to \( 0 \) value after every big event (of overlap size \( \geq M \)
where \( M=2400\) (a) and \(2048 \) (b) respectively). }{\small \par}

\section{Summary and Discussions}

\noindent In all the dynamical models of sandpile avalanches (discussed in 
section 2), we find that the growing correlations in the dynamics
of constituent elements manifest themselves as various precursors.
The number of topplings \( \Delta  \), relaxation time \( \tau  \)
and the correlation length \( \xi  \), in both BTW and Manna model,
grow and diverge following power laws as the systems approach their
respective critical points \( h_{c} \) from below (see Figs. 2 and 3): 
\( \Delta  \)
\( \sim (h_{c}-h_{av}) \)\( ^{-\delta } \), \( \tau \sim (h_{c}-h_{av})^
{-\gamma } \)
and \( \xi  \) \( \sim  \) \( (h_{c}-h_{av})^{-\nu } \). For two
dimensional systems, we find numerically here \( \delta \simeq 2.0 \),
\( \gamma \simeq 1.2 \) and \( \nu \simeq 1.0 \) for both BTW and
Manna model.
Basically, we studied the behavior
for \( h_{av}<h_{c} \), the precursor behavior, where \( \xi  \)
is necessarily finite. As we add here the tiny pulse at some central
site of a relatively large system, the boundary effect can not be
really felt because of the smallness of \( \xi  \) compared to \( L \)
for most values of \( h_{av} \). This explains the lack of finite
size effect in our precursor studies.

For the global failure or fracture in fiber bundle models (discussed in 
section 3), we find that the breakdown susceptibility
\( \chi  \) (giving the increment in the number of broken fibers
for an infinitesimal increment of load on the bundle) and the corresponding
relaxation time \( \tau  \) (required for the bundle to stabilise,
after successive failures of the fibers), both diverge as the external
load or stress approaches its global failure point \( \sigma _{c} \)
from below: 
\( \tau  \) \( \sim  \) \( (\sigma _{c} \) \( - \)\( \sigma _{av})^{-\alpha} \) 
and \( \chi  \) \( \sim  \) \( (\sigma _{c} \) \( - \)\( \sigma _{av})^{-\eta} 
\); \( \alpha=\eta=1/2 \).
These results are of course analytically derived
 for different  fiber strength distributions and the universality class 
of the model has been confirmed. 

If one Cantor set moves uniformly over another, the overlap
between the two fractals change quasi-randomly with time (see eg.,
Fig. 8). These time series are taken as model seismic activity variations 
in such two-fractal overlap model of earthquake. The overlap size 
distribution was argued \cite{BS99} and
shown \cite{SBP02} to follow power law decay. We showed numerically 
(see section 4) that if
one fixes a magnitude \( M \) of the overlap sizes \( m \), so that
overlaps with \( m\geq M \) are called `events' (or earthquake),
then the cumulative overlap \( Q_{i}(t)=\int ^{t}_{t_{i-1}}mdt \),
\(t\) \( \leq t_{i} \),
(where two successive events of \( m\geq M \) occur at times \( t_{i-1} \)
and \( t_{i} \)) grows linearly with time up to some discrete quanta
\( Q_{i}(t_{i})\cong lQ_{0} \), where \( Q_{0} \) is the minimal overlap
quantum, dependent on \( M \) and \( n \) (the generation number).
Here \( l \) is an integer (see Figs. 10, 11) \cite{Isr-03}. Although 
our results here are for regular fractals
of finite generation \( n \), the observed discretisation of the
overlap cumulant \( Q_{i} \)  with the time limit set by \( n \), is a robust
feature and can be seen for larger time series for larger generation
number \( n \). This part of the model study therefore indicates that one can
note the growth of the cumulant seismic response \( Q_{i} (t) \),
rather than the seismic event strength \( m(t) \), and anticipate some big
events as the response reaches the discrete levels \( lQ_{0} \), specific to 
the series of events.

Some prior knowledge of the precursors and their precise behavior (power laws 
etc.), as discussed here for three different kinds of failure models of 
cooperatively interacting dynamical systems which are prone to major failures, 
are therefore plausible.  
These precursor responses should help estimating
the location of the global failure or critical point using
proper extrapolation of the above quantities, which are available
long before the failure occurs. The usefulness of such precursors
can hardly be overemphasized.

\vskip.2in
\begin{acknowledgments}
\noindent We are grateful to our collaborators P. Bhattacharyya, P. 
Choudhuri and P. Ray for useful discussions.
\end{acknowledgments}

\newpage
\end{document}